\documentclass[preprint2]{aastex63}

\def\bfc{}

\def\kms{km~s$^{-1}$}
\def\cm3{cm$^{-3}$}

\def\aap{A\&A}
\def\hi{\ion{H}{1}}

\usepackage{graphicx}
\usepackage{epsf}
\usepackage{natbib}

\newcommand{\simgt}{\ga}
\newcommand{\simlt}{\lower.5ex\hbox{$\; \buildrel < \over \sim \;$}}

\providecommand{\sorthelp}[1]{}

\begin{document}

\title{Shocks and Molecules in Diffuse Interstellar Cloud Pairs}

\shorttitle{Shocked Cloud Pairs}

\author[0000-0001-8362-4094]{William T. Reach}
\affil{Universities Space Research Association, MS 232-11, Moffett Field, CA 94035, USA}
\email{wreach@sofia.usra.edu}

\author{Carl Heiles}
\affil{Astronomy Department, University of California, Berkeley, CA 94720, USA}

\begin{abstract}
The diffuse interstellar medium is dynamic, and its chemistry and evolution is determined by shock fronts 
as well as photodissociation. Shocks are implied by the supersonic motions and velocity dispersion often 
statistically called `turbulence'.  We compare models of magnetohydrodynamic (MHD) shocks, with speeds typical of cloud motions
through the ISM (3--25 \kms) and densities typical of cold neutral gas ($\sim 10^2$ \cm3), to
archival observations of the \ion{H}{1} 21-cm line for  gas kinematics, far-infrared emission for  dust mass, and
mid-infrared emission for high-resolution morphology, to identify shock fronts in three high-latitude
clouds pairs with masses of order 50 $M_\odot$.
The clouds have `heads' with extended `tails', and high-resolution images
show arcs on the leading edges of the `heads' that could be individual shocks. 
The \ion{H}{1} shows higher-velocity gas at the leading edges due to shock-accelerated material.
For two cloud pairs,  one cloud
has an active 
shock indicated by broad and offset \ion{H}{1}, while the other cloud has already been shocked
and is predominantly `CO-dark' H$_2$. 
Two-dimensional MHD simulations for shocks parallel to the magnetic field for pairs of clouds show a remarkable similarity to observed cloud features, including merged `tails' due to aligned flow and magnetic field, which leads to
lateral confinement downstream.
A parallel alignment between magnetic field and gas flow may lead to formation of 
small molecular clouds.
\end{abstract}

%\keywords{dust, ISM: abundances, ISM: atoms, ISM: clouds, ISM: general,  ISM: molecules}

\section{Introduction}

The interstellar medium (ISM) within $\sim 200$ pc of the Sun is pervaded by diffuse, low-density gas at temperatures of thousands of K (the so-called `warm neutral medium', or WNM)
and denser concentrations of gas at temperatures of order 100 K (the so-called `cold neutral medium', or CNM); 
for reviews see \citet{kulkarniheiles} and \citet{dickeylockman}.
These phases are often considered to be in pressure
balance, and a two-phase solution that can match the approximate average properties of
CNM ($T\sim 80$ K) and WNM ($T\sim 8000$ K) exists for a narrow range of thermal pressures \citep{field69,wolfire03}. 
But is the ISM really in equilibrium? Evidence suggests the contrary.
Static models cannot explain why 48\% of the atomic gas
is in the `unstable' temperature range between the CNM and WNM,  calling into
question whether there really are two distinct phases in the first place \citep{heilestroland03b}.
The prevalence of diffuse gas in the intermediate temperature range 250--1000 K was confirmed in 
a large, recent survey where 51\% of the lines of sight detected in absorption, 
amounting to 20\% of the total gas mass, comprises
an `unstable neutral medium' (UNM) \citep{murray18}.

The ISM is dynamic, being subjected to shock waves from supernovae: with a supernova occurring
every $\sim 50$ yr in the
Milky Way \citep{snrateref}, any point in the disk can be expected to be within 100 pc of a 
supernova every 10 Myr.
High-temperature ($\sim 10^6$ K) gas pervades large volumes of the ISM,
forming a `hot ionized medium' \citep[HIM; see][]{mckeeostriker}, effectively the interior of single and multiple supernova remnants.
Even without new supernova shocks, the gas is already so highly agitated that shocks will arise from CNM-WNM 
and CNM-CNM interactions, because motions of clouds are supersonic.
Cloud motions relative to surrounding gas are readily
evident from wide-area \hi\ surveys \citep{heileshabing}, with the most-common `low-velocity' clouds being within ~30 \kms\ of
the local standard of rest (LSR), `intermediate-velocity' clouds being up to 90 \kms\ \citep{magnani10}, and `high-velocity' clouds at even
higher velocities \citep{wakker97}. 
The sound speed is approximately 1 \kms\ for the CNM and 10 \kms\ for the WNM. 
If we focus upon well-defined (smaller than a few pc) concentrations of material in the local ISM with column densities $>10^{20}$ cm$^{-2}$, their densities are high enough that they will be considered
 CNM, and typical cloud velocities relative to the local standard of rest are supersonic. Within low-velocity gas, cloud motions through intercloud gas will drive shocks with
Mach numbers of 2--20.
If an intermediate-velocity cloud interacts with low-velocity gas, shocks with Mach number 
20--90 will result. And if a high-velocity cloud interacts with low-velocity gas, strong shocks with Mach numbers of 
100 and higher will result.

This simple realization means that all  interstellar `clouds' will contain shock fronts either
at their surface, possibly defining what we consider to be the `edge' of a cloud, or throughout older clouds. 
This point has not eluded astronomers in the past, and \citet{cox79} inferred that most interstellar clouds should contain at least 1 shock. 
The average Mach number in the cold, diffuse ISM excited by supernovae is estimated to be
$3.7$ \citep[see eq. 39a of][]{heilestroland05} to $\sim 2$ \citep{liostriker15}.
The very local ISM may be involved in an expanding shell from nearby OB stars, which is 
even detected as interstellar dust moving through the solar system \citep{frisch99}.
While it is tempting to rely on static, equilibrium models dominated primarily by photodissociation and radiative cooling,
nature is significantly more complicated and requires a wider vision that can also include the `unstable' \ion{H}{1} temperatures, 
the prevalence of `hot' X-ray gas, and supersonic motions. 
{\bfc While cold, atomic clouds are readily evident in 21-cm \ion{H}{1} surveys, the medium in
which they travel is not as readily evident and could be a warm atomic or ionized medium 
(WNM or CNM , $T\sim 10^4$ K) or hot ionized medium (HIM, T$\sim 10^6$ K).}

In the CNM,  evidence points toward a significant
amount of molecular gas in diffuse clouds that was not known from CO surveys. Our previous work identified a sample of clouds
that we now realize are primarily composed of `CO-dark' molecular gas \citep{reach15}, a feature of
interstellar clouds that had been independently found to be present in the ISM from infrared (HRK) and $\gamma$-ray \citep{grenier05} studies.
Analyzing \ion{H}{1} surveys showed
that cold clouds, with median internal velocity dispersion of 3.2 \kms, have total mass 2.5 times the \ion{H}{1} mass alone
\citep{kalberla20}. 
The CO-dark molecular gas is a common ingredient for galaxies \citep{leroy07}, in particular dwarf galaxies \citep{,fahrion17,chevance20}.
This additional mass is not optically-thick \ion{H}{1} \citep{lee15,reach17,murray18a}. 

To understand  many
of these diverse phenomena, we consider in this paper the possibility of finding individual shock fronts in diffuse clouds. 
Small-scale structure in the ISM is clearly evident using
higher-resolution techniques as they come online. A review of `tiny' scale atomic
structure addresses scales of 0.05 pc, which can actually be resolved for nearby clouds, and which have excess pressure
\citep{stanimirovic18}. 
We  concern ourselves here with 
identifying individual shock fronts as laboratories for detailed study. The statistical description of a medium pervaded by such shocks may be supersonic turbulence, in current astronomical terminology. The shocks set the
chemistry, thermodynamics, and kinematics of the gas and are a more apt description of the interstellar
medium than static photodissociation regions \citep[PDRs, ][]{tielens85} where starlight and
extinction are the sole determinants of chemistry. 
The energy density of starlight is 140 times less than the energy density of gas motions for nominal conditions in the cold neutral ISM with density $10^2$ cm$^{-3}$ and flow velocity 10 km~s$^{-1}$, so any mechanism that taps
into at least 1\% of the kinetic energy density is more powerful than starlight. When turbulence is dissipated
in the cold ISM, it dominates chemistry, forming turbulence dissipation regions \citep[TDRs, ][]{godard09}. In denser regions, observed supersonic CO line profiles can be explained
as intermittent turbulence \citep{falgarone09}.

The effects of starlight and 
shocks are of course always combined, and distinguishing them at an observable level remains challenging.
\citet{elitzur80} stated that ``unambiguous detection of shocks in the general interstellar medium had eluded investigators.'' Significant progress in theoretical modeling has been made in the meantime, with emphasis on
understanding the line emission from high-intensity shocks driven by stellar outflows and supernova remnants 
\citep{hm89,drainemckee93,neufeld19}. 
Molecular hydrogen emission
has been detected from diffuse clouds, potentially due to shocks \citep{ingalls11}. 
The purpose of this paper is to pinpoint shock fronts within well-defined, diffuse CNM clouds and determine their
gross physical properties. We hope to enable further studies of the physical properties of the ISM, 
both observational and theoretical. 
We utilize archival data from a large-scale \ion{H}{1} survey \citep{mcclure09}
and the {\it Planck} far-infrared dust survey \citep{planckXVIIdust}, together with theoretical calculations using the Paris-Durham shock model \citep{flowerpineau03,flower15}.

\section{Predictions for interstellar shocks\label{sec:theory}}

\subsection{Analytic}

To locate shocks in specific, diffuse interstellar clouds, we first
need to know some the distinct signatures of such shocks.
For a strong shock into diffuse gas, the temperature of the gas behind the shock
reaches a maximum of
\begin{equation}
    T_{\rm post}=\frac{3}{16} \frac{\mu m_{\rm H}}{k} v_s^2
    \label{eq:tpeak}
\end{equation}
where $\mu$ is the mean particle mass in the heated gas ($\mu=1.27$ for atomic gas, with 10\% He),
$m_{\rm H}$ is the mass of a hydrogen atom, $k$ is the Boltzmann constant, and $v_s^2$ is
the shock velocity. 
%If the width of the 21-cm line is entirely due to Doppler broadening
%from thermal motions in the post-shock gas at temperature $T_{\rm post}$, then we can 
%directly associate the line width with the shock velocity. 
For shocks at speeds 5--25 \kms,
the the peak temperature behind the shock is 690--19,000 K.
The gas behind the shock cools in a time 
\begin{equation}
    t_{\rm cool} = \frac{k T}{n\Lambda},
\end{equation}
where $\Lambda$ is the cooling rate, for the range of densities and temperatures of shocked CNM clouds.
The primary coolants for CNM gas at post-shock temperatures are atomic fine-structure lines
([\ion{O}{1}] 63 $\mu$m and  [\ion{C}{2}] 158 $\mu$m), H recombination (Lyman$\alpha$ and $\beta$) and
rovibrational lines of H$_2$. 
the cooling rate increase with temperature and leads to a cooling time for CNM gas of
\begin{equation}
    t_{\rm cool} \sim 3000 \left(\frac{n}{100\, {\rm cm}^{-3}}\right)^{-1}\, {\rm yr}.
\end{equation}
The cooling times are shorter than cloud crossing times, so the
shock fronts are shorter-lived phenomena compared to clouds as a whole.
The thickness of the layer of shock-heated gas (viewed perpendicular to the plane of the shock)
can then be estimated as 
\begin{equation}
L_{\rm cool}=\frac{1}{4} v_s t_{\rm cool}\sim 
0.01  \left(\frac{v_s}{10\, {\rm km~s}^{-1}}\right) \left(\frac{n}{100\, {\rm cm}^{-3}}\right)^{-1} {\rm pc}.
\label{eq:lcool}
\end{equation}
This sets the scale length of shocks to relate to small scale structure in interstellar clouds.

\begin{figure}
    \centering
    \includegraphics[scale=.8]{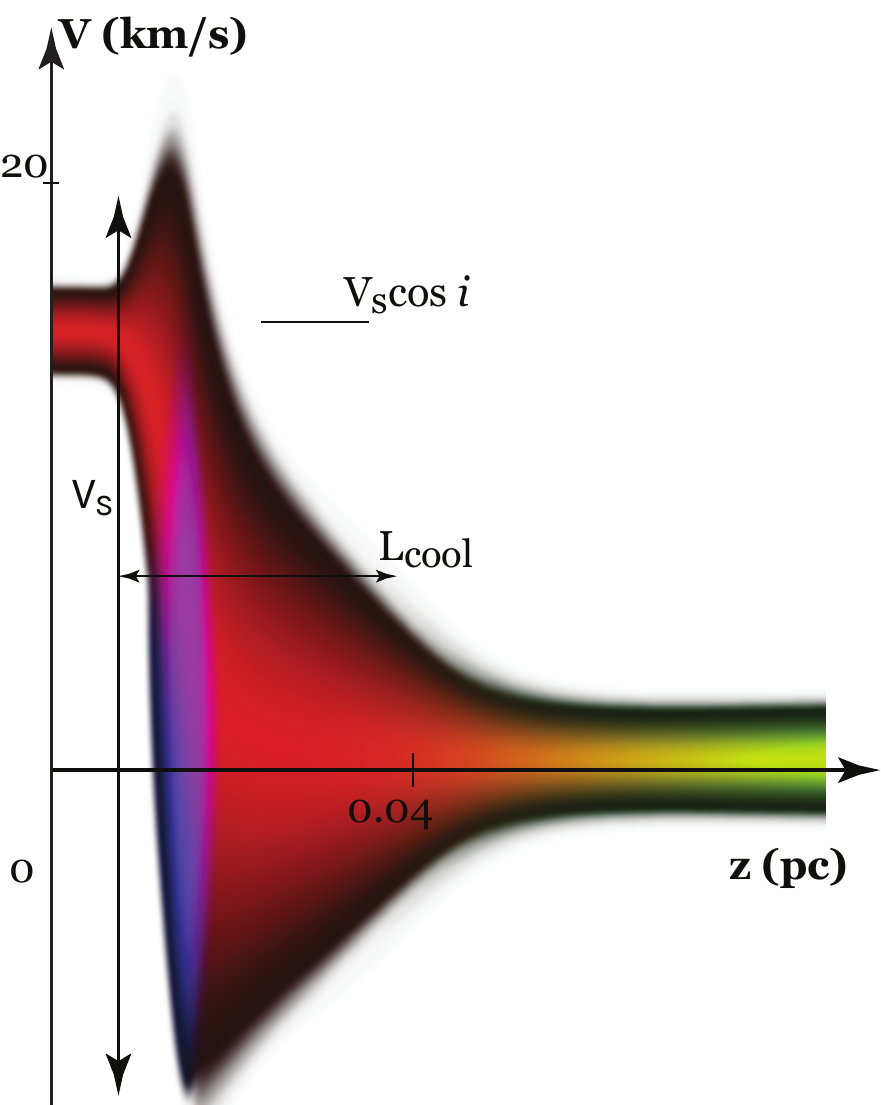}
    \caption{Theoretically predicted density behind a 20 \kms\ shock into an atomic cloud with initial density 100 \cm3. The brightness in this image is proportional to the density of \hi\ (red), H$_2$ (green) and CH$^+$ (blue). The distance behind the shock, $z$, is
    shown in a linear scale.}
    \label{fig:hih2shock}
\end{figure}

% INSERT NEW SEC FROM CARL
{\bfc 
The dynamical signature of a strong gasdynamical shock front can be
understood as follows. We describe a strong shock moving perpendicular
to the line of sight, i.e.\ with inclination angle i = 0. A strong shock
has the kinetic energy per particle dominating the thermal energy per
particle $m_H v_s^2 \gg 3 kT_0$, where subscript 0 denotes the
unshocked ambient gas.  Referring to Figure 2 as a pictorial
representation, the shock front is at the vertical double-headed
arrow. In the frame of the Universe, the shock moves into the ambient
gas from right to left with velocity $-v_s$.

It is sometimes clearer to describe velocities in the frame of the
shock, where atomic upstream gas (region 0, depicted in red, just to the
left of the shock in Figure 2) flows into the (stationary) shock from
the left, moving towards the right with velocity $v_s$.  Just behind the
shock (region 1, just to the right of the shock in Figure 2), the gas is
hot (per eq.\ 1), making the blue region in Fig. 2, and it slows to
velocity $v_s/4$.  The gas remains atomic just behind the shock because
of the short time scale. As the gas moves downstream from the shock
towards the right, it cools over the distance $L_{cool}$ (returning to
the color red on Figure 2), gets denser, and slows towards zero
velocity. In this cold, dense gas, molecules begin to form, as indicated
by the color transitioning from red to green; this defines region 2.

When a magnetic field is present we have magnetohydynamical (MHD) shocks
instead of gasdynamical shocks. The properties of MHD shocks are
determined by the relative orientations between the shock front, the
upstream magnetic field $B_0$, and the upstream velocity $V_0$; all of
these can be different. We restrict our description to strong shocks, of
which we distinguish three types: parallel shocks, perpendicular shocks,
and oblique socks.

The first two are relatively simple. For a parallel MHD shock, the
upstream magnetic field $B_0$ is aligned with the shock velocity $v_0$,
so the field has no dynamical effect: B doesn't change across the shock
front and the behavior is just like a gasdynamical shock; a strong shock
has $v_0 \gg v_{sound}$.  For a perpendicular shock, $B_0$ and $v_0$ are
perpendicular so the upstream gas drags the magnetic field lines along
with it. A strong shock has $v_0 \gg v_{Alfven}$. Flux freezing makes 
$n \propto B_{\perp}$ along the flow; for a strong shock, both $n$ and
$B_{\perp}$ increase by a factor of 4 just across the shock.

The oblique case is general and includes the parallel and perpendicular
cases \citep[\S 7.21 of ][]{fitzpatrick14}. 
For oblique MHD shocks it is convenient to convert to a frame in
which not only is the shock is stationary, but also the
upstream velocity $v_0$ and field $B_0$ are parallel; in Figure 2, this
is accomplished by adding the appropriate vertical velocity to $v_s$. In
this frame, the parallelism between $v_0$ and $B_0$ makes the upstream
(region 0) look just like the parallel MHD shock described above.
Indeed, $B_{||}$ is unchanged across the shock, i.e.\
$B_{||,1}=B_{||,0}$. But $B_{\perp,0}$ {\it does} change across the shock,
thus generating a component $B_{\perp,1}$. This non-intuitive behavior is
called a 'switch-on' shock.

For all three cases, the gas radiates thermal energy and becomes denser
as it moves downstream from region 1 toward region 2. In this flow, the
total post-shock pressure $P_{tot}=(P_{mag} +P_{th})$ stays roughly
constant. The thermal pressure $P_{th} = 3nkT/2$, so as the gas cools
the density increases with decreasing temperature; flux freezing makes
$B$ increase, too. The magnetic pressure $P_{mag}=B^2/8\pi$, so as the
temperature drops $P_{mag} \propto T^{2}$. If the field is strong
enough, the magnetic pressure will come to dominate the thermal pressure
at some particular post-shock temperature; in practice for the atomic
ISM, this does, in fact, occur.  Downstream from where this happens, the
total pressure is dominated by magnetic pressure and the density $n$ no
longer increases with decreasing temperature. At that point in the flow,
$B_{\perp}$ has increased while $B_{||}$ has remained constant, so downstream
from the shock the field lines become rotated towards being
perpendicular to the shock normal (i.e., parallel to the plane of the
shock) by an amount depending on the density ratio $(n_2/n_1)$.
}

\subsection{Computational\label{sec:computational}}

For a  detailed prediction of the properties of a diffuse cloud shock, we utilized the
Paris-Durham shock model \citep{flowerpineau03,lesaffre13, flower15} to calculate the structure and chemistry for C-type
shocks of velocity 5--20 \kms\ into gas with densities 10--1000 \cm3. 
The model accounts for gas chemistry and ion-neutral separation in the shocked gas.
An important caveat is that these are 1-dimensional models, so they do not include the diverse effects of
3-dimensional reality. We discuss comparison to 2-dimensional MHD simulations in \S\ref{sec:hydro} below.

The initial magnetic field was assumed to have strength $B=b n_0^{0.5}$~$\mu$G, where $n_0$ is the pre-shock density in \cm3, with the field orientation perpendicular to the shock velocity. With $b=1$, for a 100 cm$^{-3}$ CNM cloud, this scaling yields a magnetic field of 
10 $\mu$G. 
The actual magnetic field strength for CNM clouds of the size we are studying is not well known.
A median magnetic field strength of 6 $\mu$G was derived from 21-cm Zeeman surveys and also from synchrotron emission
\citep[see review by][]{heilescrutcher05}. The median field refers to randomly-selected lines of sight toward
radio sources. Those lines of sight are unlikely to cross the center of CNM clouds, which we show below are
highly structured, so the randomly-selected lines of sight are likely to sample regions with lower density (and magnetic field) than are sampled by 
our directed studies of cloud peaks. 

Initial conditions of the pre-shock cloud were set for a photodissociation region with the
interstellar radiation field (ISRF) at the Solar Circle \citep{mmp83}. Cosmic rays permeate
the preshock and shocked gas with constant ionization rate $\zeta=5\times 10^{-16}$ s$^{-1}$;
this cosmic ray rate is about a factor of 10 higher than used historically, inferred from relatively
recent observations of H$_3^+$ 
\citep{indriolo12} and OH$^+$ \citep{bacalla19}.

Table~\ref{modtab} summarizes the shock models.
To measure properties of the shock-heated region, 
we somewhat arbitrarily defined the cooling layer as where the temperature was 20\% higher than the final temperature.
The primary coolants in the shock-heated region are listed for each model, with the [\ion{O}{1}] 63 $\mu$m line being the primary coolant for much
of the explored parameter space, with significant contributions by the [\ion{C}{2}] 158 $\mu$m line from the slowest, low-density shocks, and 
contributions by H$_2$ from the faster shocks.
The model cooling lengths from our analytic estimate in Eq.~\ref{eq:lcool} are within
a factor of 2 of the detailed model simulations over the range of density and shock velocity considered here.

\def\tnm{\tablenotemark}
\begin{deluxetable*}{rccccl}
\tabletypesize{\scriptsize} 
\tablewidth{0pt}
%\tablenum{text}
\tablecolumns{6}
\tablecaption{Summary properties of diffuse cloud shock models\label{modtab}} 
\tablehead{
\colhead{$v_s$}  & \colhead{Compression} & \colhead{$T_{\rm max}$} &
\colhead{$N_{\rm sh}$} & \colhead{$L_{\rm cool}$}& \colhead{Coolant}\\
\colhead{(km~s$^{-1}$)}   & \colhead{($n_{\rm max}/n_{\rm min}$)} & \colhead{(K)} &
 \colhead{($10^{20}$ cm$^{-2}$)} & \colhead{(pc)} & }
\startdata
\cutinhead{$n_0=10$ \cm3, $T_0=208$ K}
  5  &  3.3 &   550   &  0.04 & 0.053 & C$^+$, O\\
  7  &  4.7 &   955   &  0.12 & 0.10 & O, C$^+$\\
 10  &  7.1 &   2500  &  0.12 & 0.12 & O, C$^+$\\
 12  &  8.7 &   3800  &  0.16 & 0.15 & O\\ 
 15  &  11  &   6100  &  0.23 & 0.19 &  O\\
 17  &  13  &   7980  &  0.29 & 0.23 &  O\\
 20  &  14  &  10,800 &  0.41 & 0.31 & O   \\
\cutinhead{$n_0=100$ \cm3, $T_0=55$ K}
  5  &  3.2 &   310   &   0.08 & 0.009 & C$^+$, O \\
  7  &  4.8 &   930   &  0.14  & 0.012 & O, C$^+$\\ 
 10  &  6.8 &   2400  &  0.23 & 0.016  & O  \\
 12  &  8.3 &   3600  &  0.31 & 0.017  & O \\
 15  &  11  &   5900  &  0.34 & 0.017  & O, H$_2$   \\
 17  &  13  &   8000  & 0.27  & 0.022  & O, H$_2$\\
 20  &  15  &  11,000 & 0.52  & 0.034  & O, H$_2$\\
\cutinhead{$n_0=1000$ \cm3, $T_0=42$ K}
  5  &  3.3 &   310   &  0.18 & 0.002  & O\\
  7  &  4.7 &   930   &  0.24 & 0.002  & O\\ 
 10  &  6.7 &  2400   &  0.41 & 0.003  & O\\
 12  & 8.1  &  3600   &  0.49 & 0.003 & O\\
 15  &  11  &  5960   &  0.59 & 0.003  & O, H$_2$\\
 17  &  13  &  8000   &  0.44 & 0.004 & O, H$_2$\\
 20  &  15  & 11,000  &  0.76 & 0.006  & O, H$_2$
\enddata
\end{deluxetable*}

\def\extra{    % version ordered by shock velocity
   5  &    10 & 208  &  3.3 &   550   &   0.04 & 0.053 & C$^+$, O\\
   5  &   100 &  55  &  3.2 &   310   &   0.08 & 0.009 & C$^+$, O \\
   5  &  1000 &  42  &  3.3 &   310   &  0.18   & 0.002  & O\\
  10  &    10 & 208  &  7.1 &   2500   &  0.12 & 0.12 & O, C$^+$\\
  10  &   100 &  55  &  6.8 &   2400   &  0.23 & 0.016 & O  \\
  10  &  1000 &  42  &  6.7 &   2400  &  0.41    & 0.003 & O\\
  15  &    10 & 208  &  11  &   6100   &  0.23 & 0.19 &  O\\
  15  &   100 &  55  &  11  &   5900   &  0.34 & 0.017 & H$_2$, O   \\
  15  &  1000 &  42  &  11  &   5960   &  0.59 & 0.003 & O, H$_2$\\
  20  &    10 & 208  &  14  &  10,800   & 0.10  & 0.038 & H$_2$, O   \\
  20  &   100 &  53  &  15  &  11,000   & 0.52  & 0.034   & O, H$_2$\\
  20  &  1000 &  42  &  15  &  11,000  &  0.76  & 0.006  & O, H$_2$\\
}

\begin{figure*}
\plotone{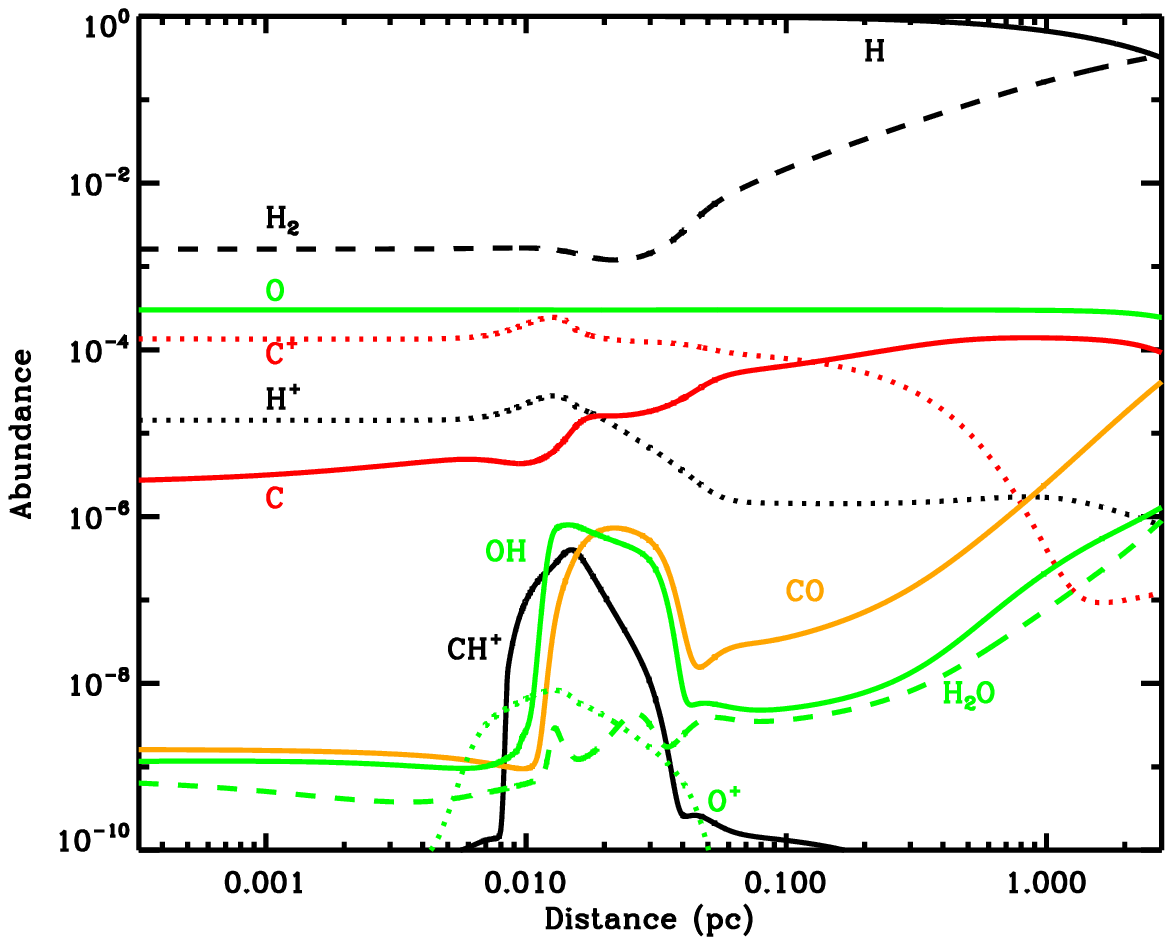}
\caption{Abundances behind a shock front of 20 \kms\ speed into gas with density 100 \cm3, for the case where the magnetic field is perpendicular to the shock velocity. The shock-heated layer is at 0.01 to 0.04 pc, where some species rapidly increase in abundance. 
Most neutral molecules increase in abundance in the cooler layer behind the shock, primarily due to extinction and self-shielding.
\label{shock-abund}}
\end{figure*}

The rapid changes in gas properties occur in a transition layer that is $\sim 0.02$ pc thick for CNM conditions and a nominal, 10 \kms\ shock.
Figure~\ref{shock-abund} shows the abundances versus distance behind the nominal shock. 
The H chemistry, which is of primary importance for this project, shows the shocked layer 
primarily atomic, with H$_2$ formation only at the highest column densities where extinction
slows photodissociation.
The O abundance remains constant throughout the shock, with is important because of its
role in cooling the shock-heated gas.
The abundances behind the shock reveal certain molecules characteristic of non-equilibrium chemistry, like CH$^+$ and O$^+$,
which are produced in and largely confined to the heated layer just behind the shock front.
Many molecular species, like H$_2$ and H$_2$O, gradually increase in the cooling gas, but they are primarily
increasing because the column density and extinction increase.
The carbon chemistry is driven primarily by ISRF ionization and has the
structure of a photodissociation region. Oxygen remains primarily atomic,
with O-bearing molecules and ions more than an order of magnitude lower in abundance.
Some molecules, like OH and CO are produced both in the shock and in the cooling gas.

At each distance behind the shock, and for each gas species, we can define an emissivity 
as a function of velocity by taking the modeled abundance of the species and spreading 
it as a Gaussian with thermal dispersion 
\begin{equation}
\sigma(p,z)=\sqrt{\frac{k T(p,z)}{\mu_p m_H}},  
\label{eq:sigma}
\end{equation}
where $p$ is the species 
(H, H$_2$, and CH$^+$ for this Figure) and $z$ is the distance behind the shock. 
The central velocity for each species at each distance behind the shock is set to neutral gas velocity
(H and H$_2$) or ion velocity (CH$^+$).
This procedure yields the velocity distribution  for each species,
\begin{equation}
\frac{{\rm d}n}{{\rm d}v}  (p, z, v) = \frac{n(z)
e^{-\left[v-\left(v_p(z)-v_s\right){\rm cos}i\right]^2/2\sigma^2}}
{\sqrt{2\pi}\sigma} 
\label{eq:emiss}
\end{equation}
 where $\sigma$ is the thermal dispersion per equation~\ref{eq:sigma} and $v_p(z)$ is the velocity at distance $z$ behind the shock appropriate to each species (neutral for H and H$_2$ and
ionized for CH$^+$).

Figure~\ref{fig:hih2shock} shows the  velocity distribution of  \ion{H}{1}, H$_2$, and CH$^+$ density
for  a 20 \kms\ shock into gas with initial density 100 cm$^{-3}$.
The relation between the shock velocity projected by inclination to the line of sight, speed of the
shocked and cooled gas, and the cooling layer, as introduced in the analytic section above, are indicated
on the Figure. The evolution and velocity distribution of the three species is distinct:

\begin{itemize}
\item The abundance of CH$^+$, chosen for illustration as a shock-only tracer,
peaks just behind the shock front then rapidly drops as the gas cools. One signature of 
a diffuse shock is CH$+$ emission in a thin layer and wide velocity dispersion. This
signature is worthy of future study and can be used to confirm shocks, once we locate them.

\item The atomic H is predicted to have a wide velocity profile for a line of sight centered on the shock front,
with width of order $V_S$, within a layer of size $L_{\rm cool}$, which is
approximately 0.03 pc for the  model case shown here. The width of the \hi\ 21-cm line at the 
location of the shock front is therefore a practical indicator of the shock velocity. The shocked
gas is readily distinguished from the pre-shock gas, which as has a narrow linewidth and is centered
at zero velocity.

\item Further behind the shock front,
if it evolves long enough and the cloud is large enough, H$_2$ appears.
Other than the molecular ions that appear only in the shock front, the molecular gas tracers are
predicted to have a relatively 
narrow velocity component and to arise approximately 0.1 pc behind the shock front.
Because the H$_2$ is cold, and the first energy level of H$_2$ is 520 K above the ground state,
there is no detectable emission from H$_2$. A proxy such as a low-J CO transition could detect
the cold molecular gas, though quantitative usage of the brightness to determine the amount of 
molecular gas is problematic given the line optical depth would be unknown and the diffuse 
clouds we are studying are unlikely to be virialized. 
\end{itemize}

%\clearpage

\section{Sample of Interstellar Cloud Pairs}

To provide candidate locations for locating interstellar shocks,
we selected
isolated, 
approximately degree-sized regions of enhanced ISM column density. 
Clouds were identified from far-infrared (100 $\mu$m) dust images from {\it IRAS}, which provided 
what was at the time by far the best tracer of the structure of the diffuse interstellar medium \citep{low84}.
The clouds were clearly visible as peaks 
both in dust and atomic gas distributions and well separated from the galactic plane (where confusion along
the line of sight becomes significant).
The initial sample was selected by inspecting the {\it IRAS} 100 $\mu$m  
and Hat Creek 21-cm all-sky surveys
\citep[][hereafter HRK]{hrk88}.
Of particular interest for our study of shocks, some clouds were in pairs. While any single cloud
could have a preferred direction of uncertain origin, pairs of clouds that share morphological features indicating 
a preferred direction offer a  clearer indication of hydrodynamic effects. 
Pairs have been important in many areas of astronomy, where
studies of star clusters allow investigation of stellar evolution \citep{binney98,krumholz19} and star \citep{luhman12} and planet formation \citep{haisch01},
because stars in each cluster are at nearly the same distance from the Sun and arguably have similar ages.
A  key advantages of studying interstellar cloud pairs are that their components are at 
the same distance, will likely have the same age, and they travel through comparable
areas of the galaxy with comparable speeds.

Distances to interstellar clouds are generally not well known. 
Assuming the clouds are within the scale height of the galactic disk
$H\sim 230$ pc derived
from local \ion{H}{1} or  [\ion{C}{2}] scaled to the solar neighborhood \citep{langer14},
clouds at 45$^\circ$ latitude  have distances within 160 pc. 
To assess the distance quantitatively, we used the three-dimensional dust reddening map developed by \citet{green19}. The great advances with GAIA, Pan-STARRS, and 2MASS, having wide coverage with accurate photometry and parallactic distances makes 
it possible to measure actual distances for some interstellar features that
heretofore had to be considered unknown.
When working on the clouds in general, we will assume a distance of $100 d_{100}$ pc for clouds with no known distance, leaving $d_{100}$ as a free parameter.

The dust column density used in this project was calculated
from the far-infrared
opacity at $5'$ resolution from the {\it Planck} survey
\citep{tauber2010a,Planck2011-1.1},
specifically the `thermal dust' foreground separation
\citep{planck2013viii}.

\hi\ 21-cm line data are from
the Parkes Galactic All-Sky Survey \citep[GASS; ][]{mcclure09}. 
This survey used the 64-m Parkes antenna with its multi-beam feed.
The survey has  angular resolution 15$'$ and velocity resolution 0.8 \kms, making
them key for measuring systematic gas motions that are signatures of interstellar shocks.
We specifically use the  GASSIII release \citep{kalberla15},
which has side lobes and stray light corrected.

The mid-infrared surface brightness was taken from the
{\it Wide Field Infrared Survey Explorer (WISE)}
all-sky survey \citep{wright10}, specifically the {\it WISE} Image Atlas served through
the Infrared Science Archive.
We also used the unWISE coadds \citep{meisner14}, which provide the 
highest angular resolution ($15''$) all-sky survey of dust emission from the interstellar medium. The unWISE images
are at 12 $\mu$m wavelength and have point sources removed, making them ideal for extended emission studies of 
interstellar clouds.

\clearpage

\section{G228 and G230}

\subsection{Morphology and kinematics}

In our prior papers \citep[HRK, ][]{reach94}, the cloud G228 was listed as G228.0-28.6,
and G230 was listed as G230.1-28.4. 
%, and they are the prominent ISM clouds in Lepus.
The extinction versus distance from the Gaia-Pan-STARRS-2MASS compendium \citep{green19} 
had no stars with less than the full extinction of the clouds, while all stars
beyond 260 pc  are
essentially fully extinguished, setting an upper limit of $d_{100} < 2.6$.
The clouds therefore lie no more than 120 pc from the galactic plane,
i.e. within one scale-height of the atomic ISM.

\begin{figure}
\epsscale{1}
\plotone{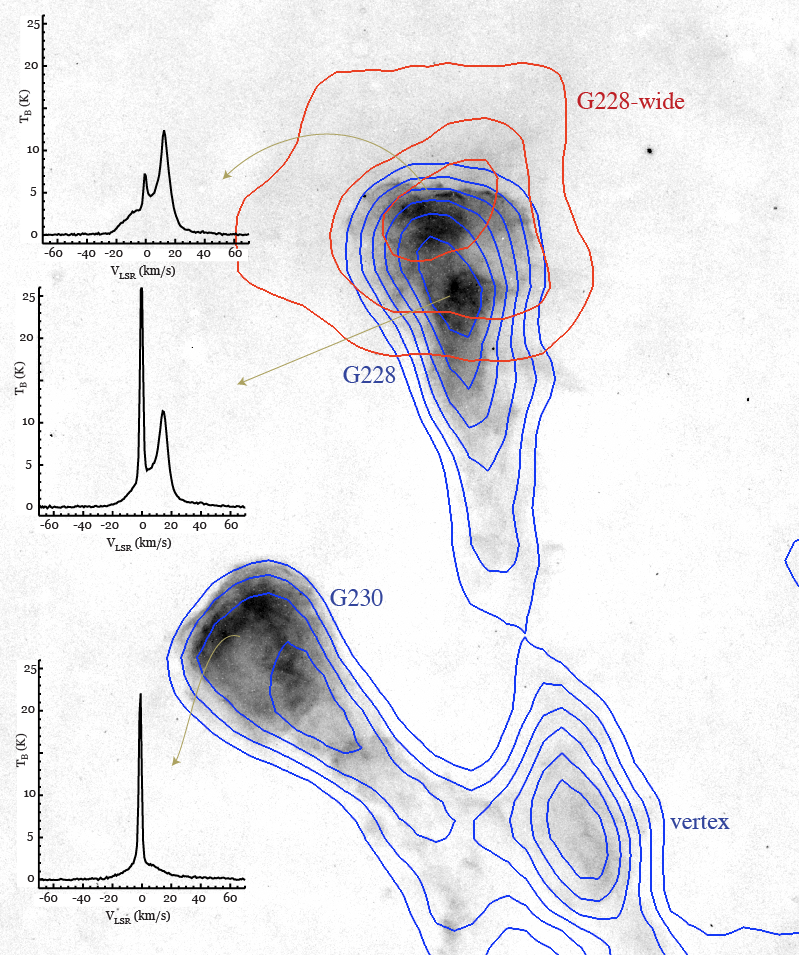}
\caption{
Comparison of the dust (un{\it WISE}) and atomic gas (GASS \hi\ 21-cm line) distributions for the clouds G228 and G230.
The dust map is in greyscale (black meaning more dust), and the atomic gas contours are overlaid. The blue contours are integrated over the narrow \hi\ component, showing the bodies of the clouds.
The red contours are the 21-cm line integrated over the wider component. 
The broader emission line is located  upstream of the head of G228 is labeled as G2280wide, 
and the `vertex' where the tails of the two clouds rejoin is labeled. Contours range from 
1--5$\times 10^{20}$ cm$^{-2}$.
The \ion{H}{1} 21-cm spectra toward three positions are shown as insets, with arrows indicating their spatial
locations near the peak of the wide-line region, the head of G228 and the head of G230.
\label{G228and230gass}}
\end{figure}

\begin{figure}
\plotone{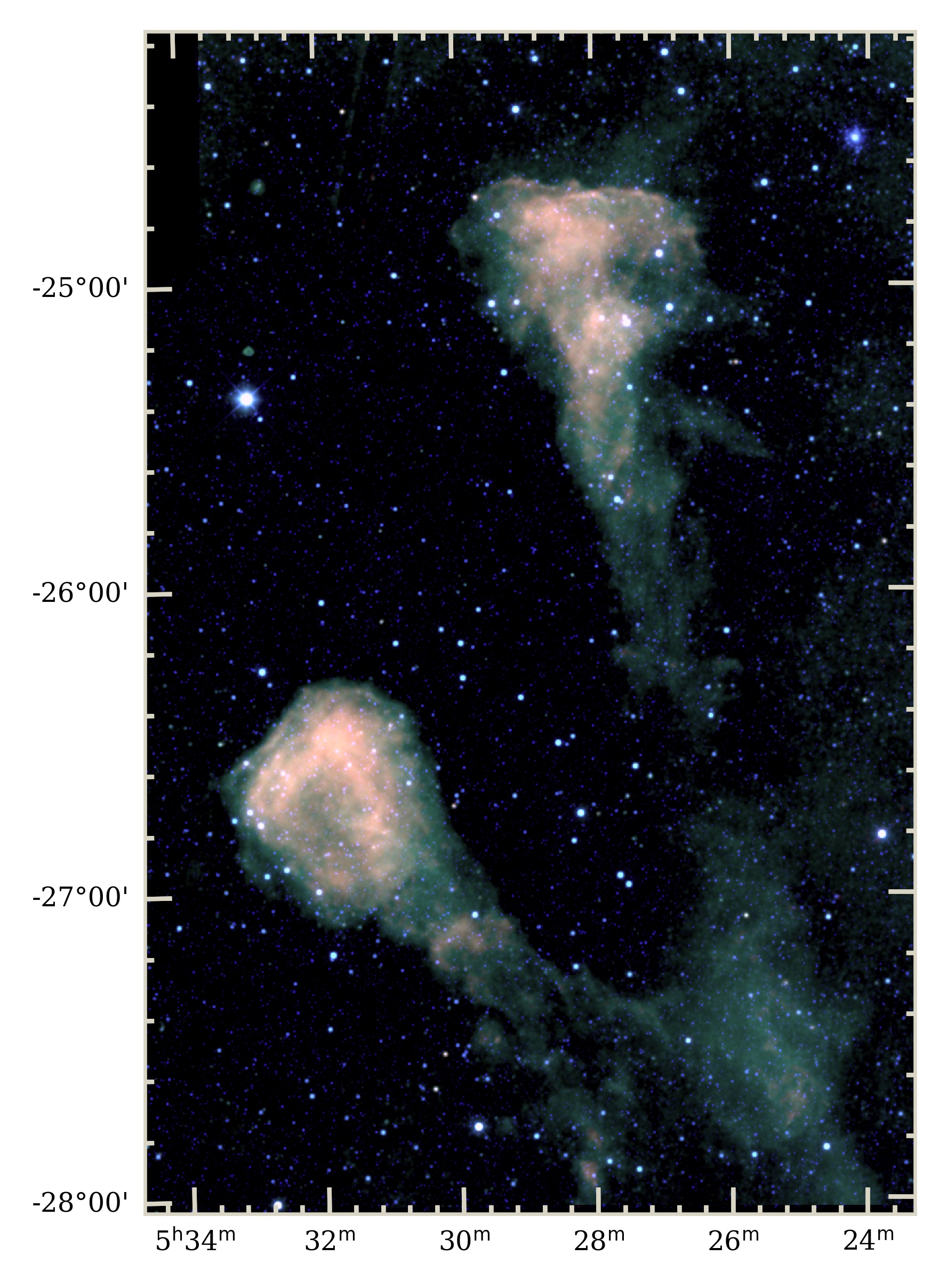}
\caption{{\it WISE} portrait of clouds G228 and G230, combining the 3.6 and 4.5 $\mu$m (blue), 12 $\mu$m (green) and 22 $\mu$m (red) images.
\label{G228and230color}}
\end{figure}

Figure~\ref{G228and230gass} shows the GASS \hi\ and {\it Planck} dust maps containing G228 and G230. 
From the Planck integrated optical depth map, assuming gas-to-dust mass ratio of 100 and optical depth (at 353 GHz) per unit H column density of 
$9.7\times 10^{-25}$ cm$^{2}$, the dust mass of the complex comprising the two clouds is $47 d_{100}^2 M_\odot$. 
Morphologically, each cloud comprises a `head-tail' structure. 
The `heads' are G228 and G230, which have masses of 23 and 14 $d_{100}^2 M_\odot$, respectively.
The volume densities of the cloud heads are both approximately $500 d_{100}^{-1}$ cm$^{-3}$ if H is atomic, or half that if the H is molecular.
These two clouds have 
radically different \hi\ properties, despite appearing similar in total column density of dust and gas. This situation was noticed in in HRK: while G230 is dominated by a very narrow (1 km~s$^{-1}$  wide) \hi\ profile, the other cloud G228 has two components: a narrow one, at approximately the same velocity as that of G230, and a broader (11.5 km~s$^{-1}$ wide) component at a distinctly different central velocity. Based on the distinct \hi\ profiles, we asserted that the clouds have different recent shock histories. The narrow component was attributed to fully-cooled gas, and the broader component to more recent or ongoing shocks, so G230 was shocked and has cooled and condensed into a molecular cloud, while G228 is still being shocked.

The {\it WISE} image (Fig.~\ref{G228and230color}) shows that G228 contains a network of filaments parallel to
an abrupt northern edge. 
The filaments can be partially discerned in deep optical images of the clouds as well \citep{stark95}.
We believe those filaments and the abrupt northern edges of the clouds are low-speed shock fronts, 
supporting our earlier speculation of the nature of these clouds.
There are also arc-like structures defining the `leading' edge of G230, where we take `leading' to mean the opposite direction
from their `tails' that are evident in Figures~\ref{G228and230gass} and ~\ref{G228and230color}.
The apparent shock fronts in the {\it WISE} image of G228 are narrow, with the sharpest ones at the leading edge of G228 
having width approximately $20''$ that is not quite resolved at the $15''$ resolution of the image. This sets an upper limit 
to the width of the heated region behind the shock, $L_{\rm cool}<0.007 d_{100}$ pc. 
For a shock velocity of 11 \kms,
the thin structure is consistent with expectations for an individual shock as long as the pre-shock density was greater 
than 50 cm$^{-3}$. Given the present density of the cloud being 10 times higher, it appears likely this condition was met.
The \hi\ 21-cm
column density at the leading edge of G228 is $4.7\times 10^{20}$ cm$^{-2}$, averaged within the $15'$ beam
of the telescope. If the line of sight depth is greater than the beam size but smaller than the cloud size, 
then the average \hi\ density is $>$ 60 $d_{100}^{-1}$ cm$^{-3}$. 
Thus it appears that
the narrow filamentary structures at the leading edge of G228 are consistent with individual shock fronts
into relatively dense CNM gas.

The shock on the leading edge of G228 appears redder in color than the rest of the cloud
in the mid-infrared (Fig.~\ref{G228and230color}, indicating relative prominence of 22 $\mu$m emission.
Some color variation could be created by different angular resolutions of the images comprising the image, but this effect
would cause sharp structures (unresolved at the longer wavelengths) to appear more blue or green, not red.
The mid-infrared color of the filaments within the clouds may be related to
their containing active shock fronts. Two possible mechanisms could explain the color difference of the apparent shocks.
One is shattering of larger grains, leading to an enhancement in 
the abundance of small grains responsible for 22 $\mu$m emission but not enhancing the shorter-wavelength emission. 
The other mechanism for the red color of the apparent shocks
is emission from the shocked gas. 
The surface brightness of the line emission would be 
$2\times 10^{-5}$ erg~s$^{-1}$~cm$^{-2}$~sr$^{-1}$ to provide the
observed brightness.
The WISE 22 $\mu$m band spans 20--26 $\mu$m. The
obvious choice is the H$_2$ (2--0) line at 28.2 $\mu$m, which has been detected from diffuse clouds and
could be potentially attributed to MHD shocks \citep{ingalls11}, but this is outside the WISE passband. 
Other possible lines include 
[\ion{Fe}{2}] at 25.99 $\mu$m, and [\ion{S}{1}] at 25.25 $\mu$m;
both lines are detected in MHD shocks from outflows
\citep{neufeld09}.

\begin{figure*}
\includegraphics[width=3in]{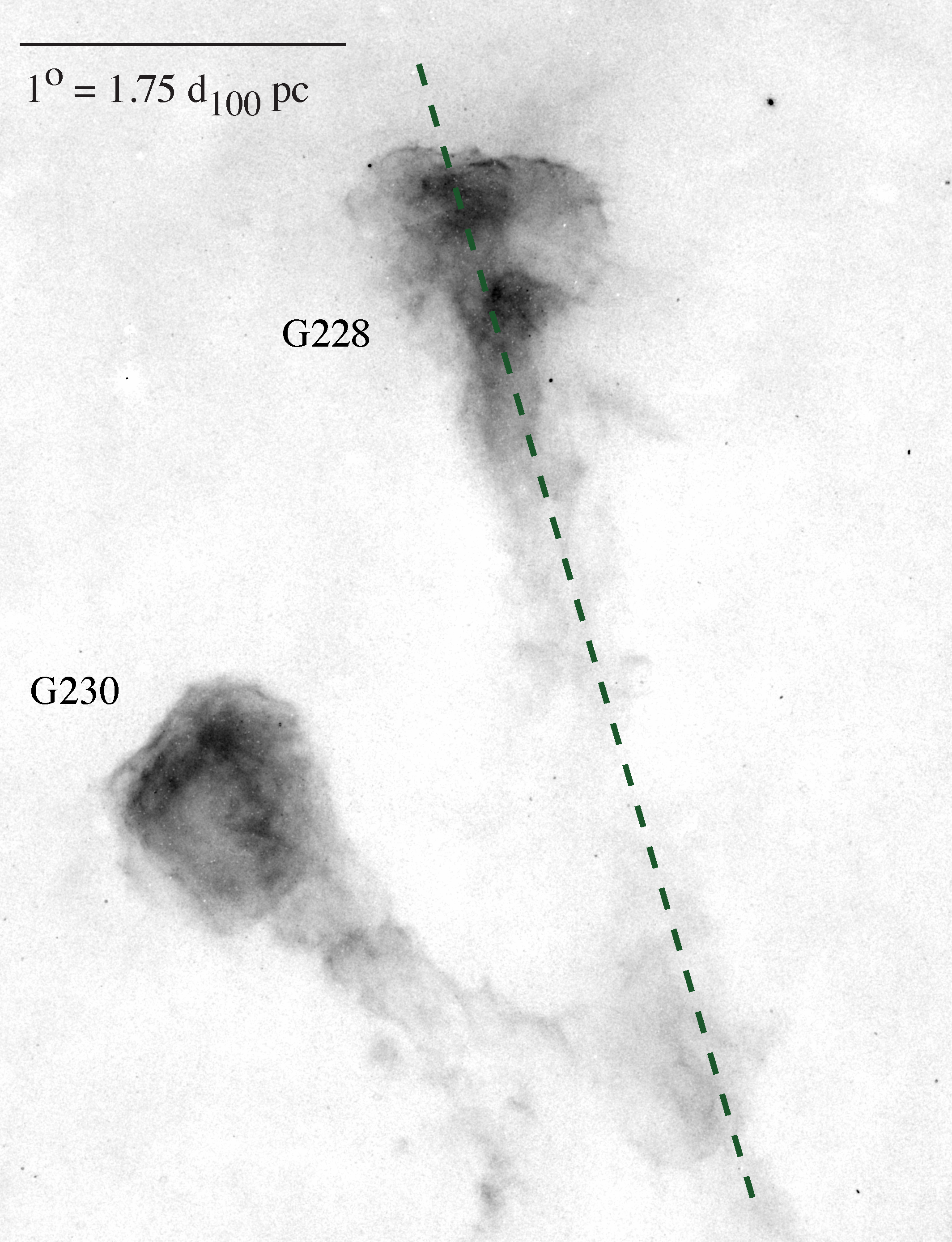}\includegraphics[width=4in]{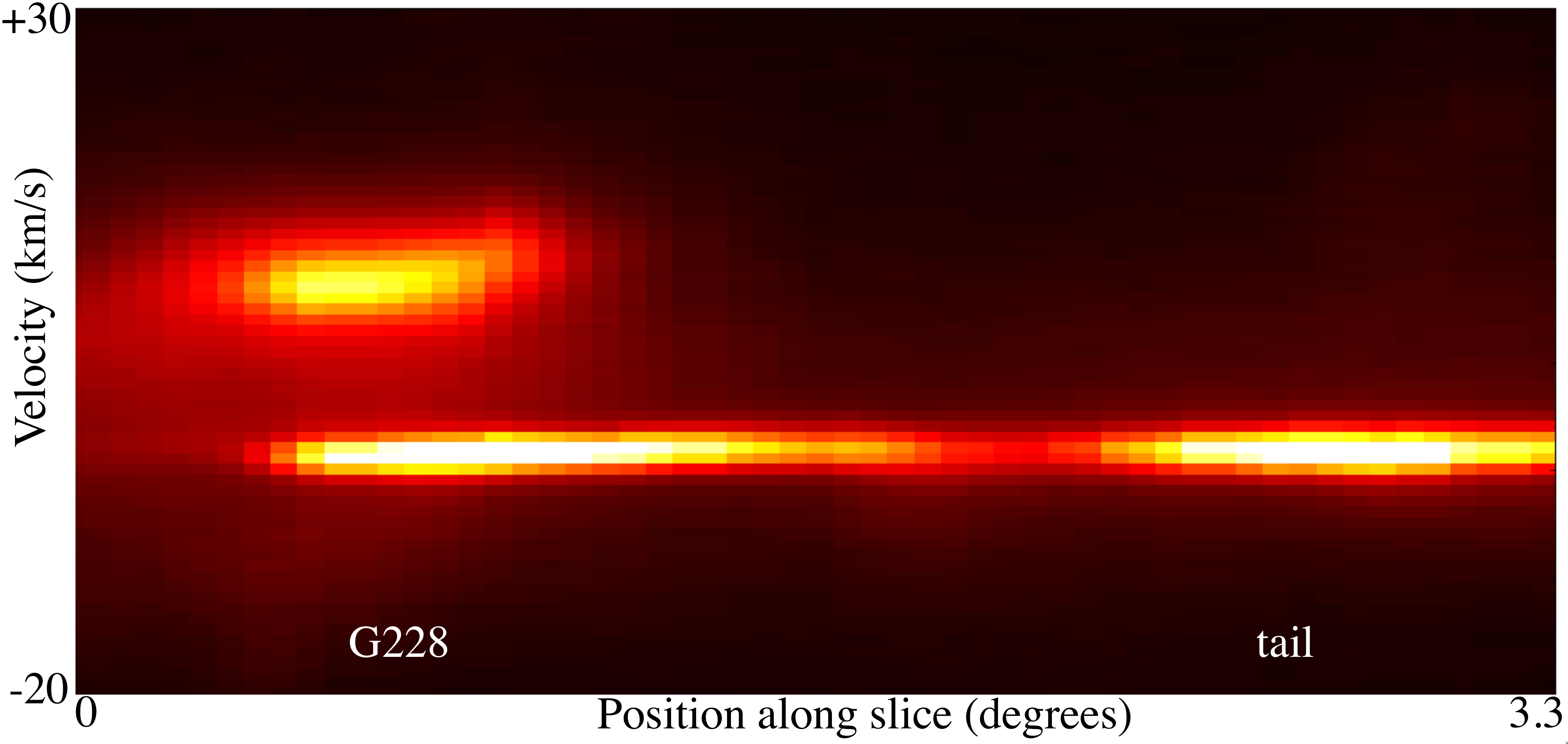}
\caption{Position-velocity slice for clouds G228 and G230. 
The left-hand panel is the unWISE mid-infrared image, with a dashed green line illustrating the path along which the slice was made. The right-hand panel shows the 21-cm line brightness temperature, with the horizontal
axis being position along the slice (starting from 0 at the top of the green line in the left-hand 
panel), and the vertical axis being the velocity (0.8 km~s$^{-1}$ pixels, spanning 50 km~s$^{-1}$).
\label{G228and230slice}}
\end{figure*}

The GASS, {\it Planck},  and {\it WISE} surveys shed significant new light on the nature and histories of the two clouds. The \hi\ profiles from our earlier paper (HRK) are confirmed, including the assertion that the two velocity components both are associated with the infrared clouds (as opposed to being  a widespread, diffuse emission). In fact, the wide component is only associated with G228, and it is somewhat spatially offset from the narrow component. 
Figure~\ref{G228and230gass} shows that the wide \hi\ component peaks on the `leading' edge of the cloud, if we envision the cloud moving through the ISM with the tenuous tail `trailing' behind it. We interpret the wider component as recently shocked gas.
Figure~\ref{G228and230slice} shows a position-velocity slice through G228, with the accelerated gas evident
as a wider spectral line, spatially coincident with the leading edge of G228, at a velocity offset from the
narrow-line emission. 
This gas should be tightly associated with the thin ridges in the WISE images, which are
likely the shock fronts. At the angular resolution of the \hi\ data, this appears to be
the case. 

\subsection{Atomic and molecular content}

\begin{figure*}
\includegraphics[width=2in]{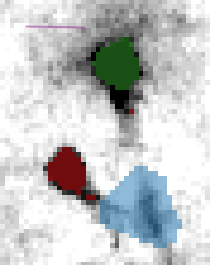}\includegraphics[width=5in]{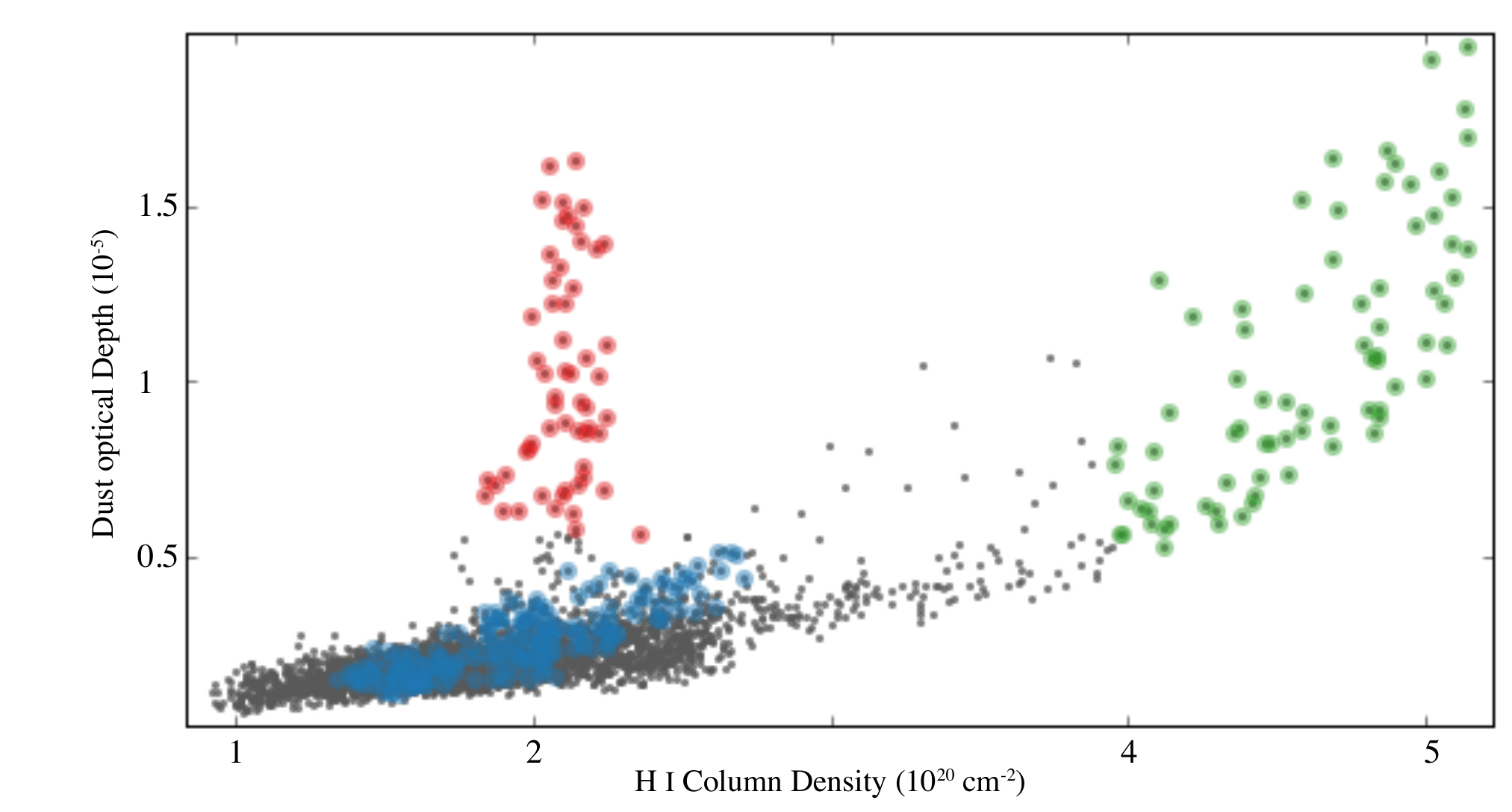}
\caption{Comparison  of dust optical depth and \hi\ column density for G228 and G230.
The purple scale bar is $1^\circ$ long (1.75 pc if the cloud is at a distance of 100 pc).
The left-hand panel shows the dust image, and the right-hand panel shows the scatter-plot.
Subsets of the data are shaded in color on the left-hand image and have colored symbols on the right-hand panel. 
In red are the points with the highest excess
of dust optical depth compared to \hi\ column density; these locations are
concentrated into cloud G230, where CO was also detected. 
In green are the points selected to have the highest \hi\ column density. 
The \hi\ peak is in G228, and not coincident with the dust peak, which is in G230.
In blue are the points selected in the southern `tail' of the  image south of G228 and G230.
This region has roughly the same dust-versus-gas trend as the rest of the image and is normal, diffuse atomic gas. This figure, and similar ones in this paper, was made using the package {\tt glue} \citep{beaumont15}.
\label{G228and230scat}}
\end{figure*}

Comparing the {\it Planck} and GASS surveys further highlights the drastic differences between the G228 and G230. Figure ~\ref{G228and230scat} shows that that G230 has a far higher amount of dust per unit atomic gas than G228. We showed earlier that both G228 and G230 contain CO in their cores, with central velocity and width similar to those of the narrow \hi\ component, and CO core brightness comparable between the two clouds \cite{reach94}. 

The {\it Planck} and GASS surveys show that the amount of dust per unit atomic gas is radically different between G228 and G230. 
If we use the gas column density from the \hi\ 21-cm line and the dust optical depth from far-infrared emission \citep[cf. eq. 4 of][]{reach15}, 
the \textit{atomic} gas/dust ratio [G/D]=84 in the regions surrounding the clouds, [G/D]=55 in G228, and [G/D]=12 in G230. 
If the total gas-to-dust ratio is 84 everywhere, then there is significant non-atomic gas in both clouds.
The amount of this `dark' gas per unit atomic gas \citep[eq. 7 of][]{reach15} is 
$f_{\rm dark}=1.5$ in G228 and $f_{\rm dark}=7$ in G230. 
There is no large-scale CO emission  from the main bodies of the clouds in the {\it Planck} CO survey, which sets a limit 
$W({\rm CO})< 7$ K~km~s$^{-1}$ on widespread CO.
But we did detect CO from the cloud cores
using ground-based telescopes \citep{reach94};
the CO brightness was $W({\rm CO})=2.1$~K~km~s$^{-1}$. 
Mapping observations showed bright CO, extended over approximately $6'\times 6'$ \citep{stark95}.
For G228, we can estimate the amount of `dark' gas that is required, in addition to the \hi, to yield a standard gas-to-dust mass ratio:   
$N_{\rm dark}=3\times 10^{20}$ cm$^{-2}$.
Combining the inferred dark gas column (assumed to be H$_2$) and CO line integral yields the so-called ``X factor'', 
$X_{\rm CO}\equiv N({\rm H}_2)/W({\rm CO}) \sim 1\times 10^{20}$ cm$^{-2}$/(K~km~s$^{-1}$), 
if the dark gas is H$_2$. 
In fact, the `dark gas' is much more widely distributed than the CO detections, so the actual values of $X_{\rm CO}$ range from the value at the core quoted above to values $>10\times 10^{20}$ cm$^{-2}$/(K~km~s$^{-1}$) in the rest of the cloud.
For G230, the bulk of the cloud mass is not atomic; if the excess gas is molecular, then its
central column density (at the 16$'$ angular resolution
of the \hi\ 21-cm line observations) is $N({\rm H}_2)\simeq 7\times 10^{20}$ cm$^{-2}$ and the inferred
$X_{\rm CO}\simeq 3\times  10^{20}$ cm$^{-2}$/(K~km~s$^{-1}$), which is reasonably similar to the value found in the Galactic Plane \citep{xfactorreview}.
Based on these numbers we find that G228 is mostly atomic, comprising approximately 52\% atomic gas,
26\% dark molecular gas, and 22\% CO-traced molecular gas. 
On the other hand G230 is predominantly molecular, with 10\% atomic gas, 80\% dark molecular gas, 
and 10\% CO-traced molecular gas.
The dust temperatures in the two clouds (17.4 K in G230 and 18.1 K in G228) and the  
diffuse medium outside the clouds
(20.5 K) are in accord with the trend of molecular content
versus temperature seen in the sample of clouds observed with Arecibo \citep[Fig. 12 of][]{reach17b}.

To search for evidence of grain processing in the shocks at the leading edge
of G228, where the wide \hi\ component and narrow apparent shock fronts are present, 
we carefully measured the brightness of the arc and the overall cloud, relative to the sky immediately adjacent,
in WISE bands 3 and 4. We find the brightness ratio between the two WISE bands is
identical within uncertainties for the arc and the overall cloud. 
Therefore, there is either no grain processing or the processing maintains the same color ratio. WISE band 3 measures the
abundance of PAH, while WISE band 4 measures the abundance of transiently heated very small grains (VSG). Maintaining a constant ratio of PAH/VSG indicates the smallest particles
have the same size distribution in the apparent shock fronts and in the main body of the cloud. The abundance ratio of small grains to big grains (BG), which maintain equilibrium temperatures, is slightly elevated for G228 and G230 compared other clouds, both for VSG and PAH \citep{reach17}.

\section{G229 and G225}

\subsection{Morphology and kinematics}

\begin{figure*}
\includegraphics[width=3in]{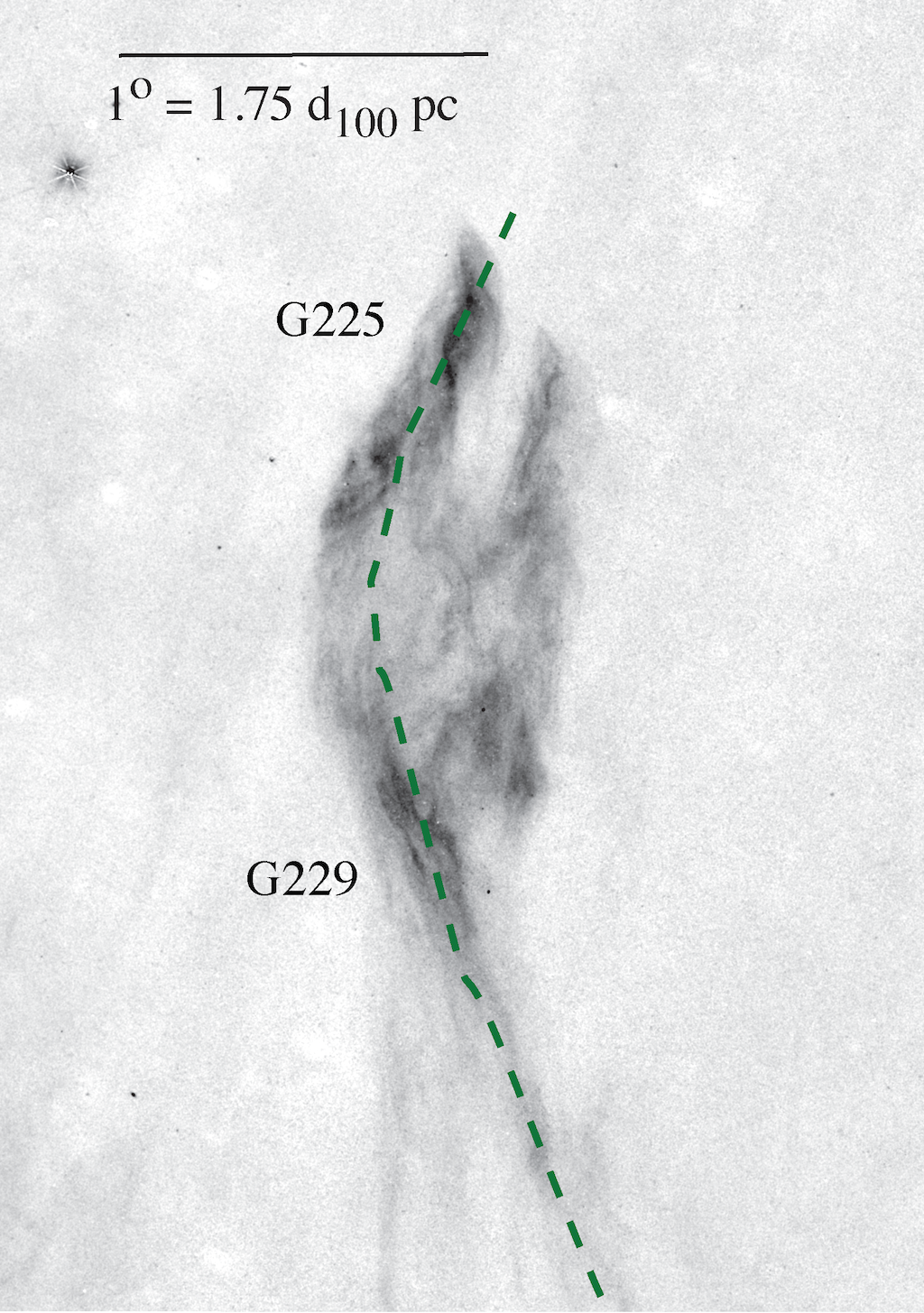}\includegraphics[width=4in]{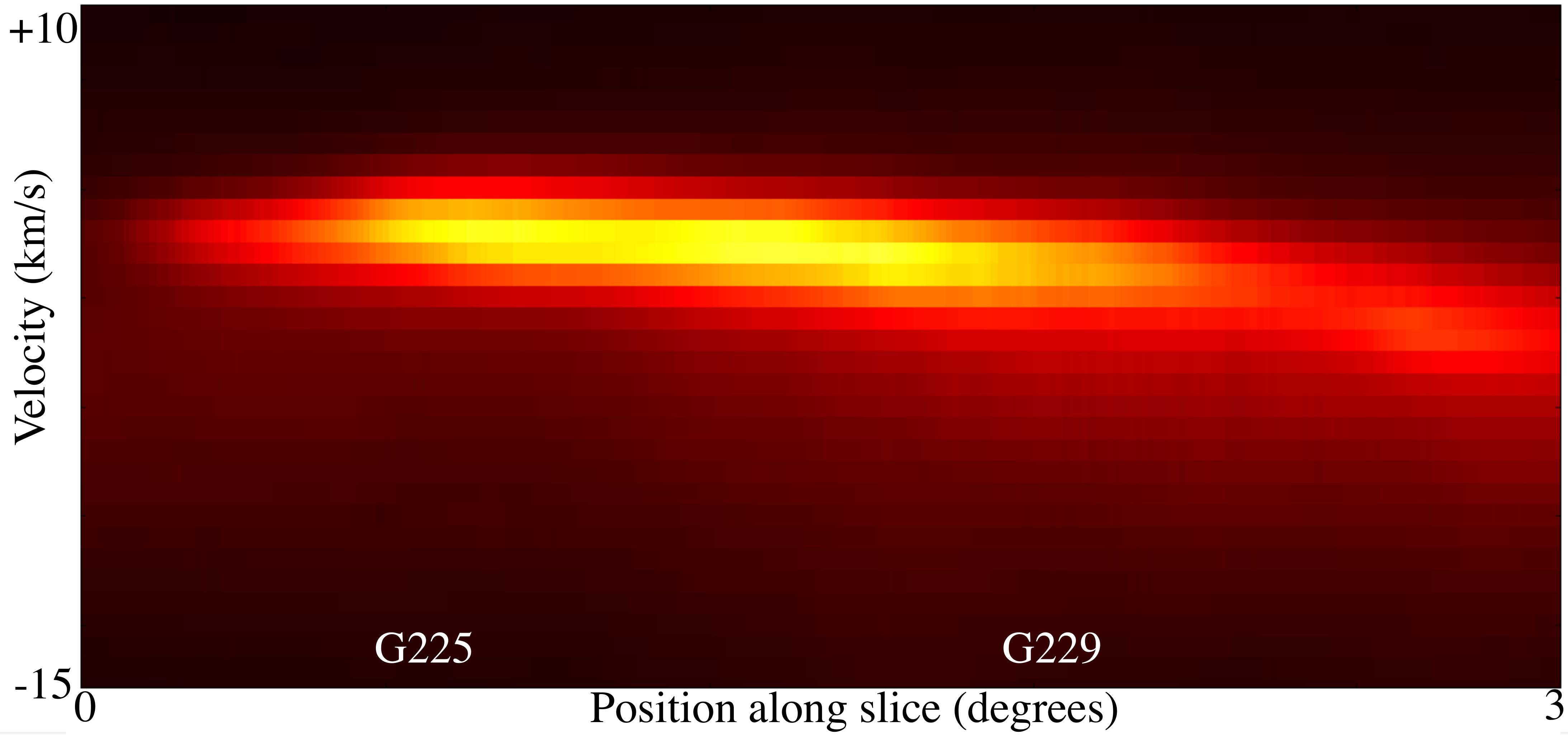}
\caption{Position-velocity slice for clouds G229 and G225. The left-hand panel is the unWISE map of the cloud, with a dashed green line illustrating the path along which the 
slice was made. The right-hand panel shows the 21-cm line brightness image, with the horizontal
axis being position along the slice and the vertical axis being the velocity. 
\label{G229and225slice}}
\end{figure*}

In our prior papers \citep[HRK, ][]{reach17}, the cloud G225 was listed as G225.6-66.4,
and the cloud G229 was listed as G229.0-66.1.
The two clouds are peaks within a complex that comprises the most prominent optical interstellar feature in the Fornax constellation as noted by \citet{paley91}.
The extinction versus distance shows that most stars in this 
direction are  fully extincted by the clouds, setting an upper limit of $d < 250$ pc. 
The Bayesian analysis of \citet{green19} for the lines of sight toward these clouds does
show a  step function at a common distance
$230\pm 25$ pc, meaning these clouds are 210 pc below the galactic midplane, approximately
one scale-height of the atomic gas.
The dust mass of the complex comprising the two clouds is $200  M_\odot$. 
Morphologically, the cloud comprises a pair of `head-tail' structures, with the tails merging
at a vertex.
One `head' is G225 which has as mass of 48 $d_{100}^2 M_\odot$, respectively.
The density in the G225 is highest, at 1000 \cm3, while the G229 portion has a density of 70 $d_{100}^{-1}$ \cm3.

Figure~\ref{G229and225slice} shows the mid-infrared image and \hi\ 21-cm position-velocity diagram. 
The cloud can be described as two bright `heads' followed by roughly parallel tails. 
The northern head is G225, which also appears in the IRAS point source catalog as 
IRAS02365-2950, while the southern head is IRAS02356-2959 \citep{irasexpsupp}. Like
many other 100 $\mu$m-only sources in the IRAS catalogs, these are not truly compact sources 
and are peaks of interstellar clouds \citep{reach93}; 
when viewed at higher resolution in Figure~\ref{G229and225slice}, they  comprise extended 
structure---in particular, bright, arced filaments extending along the direction connecting to the more tenuous tails. The cloud was described similarly by \citet{odenwald88} as ``resolved into two nuclei, each having its own sinuous tail extending $\sim 14^\circ$ to the south,'' which is even further than shown in Figure~\ref{G229and225slice}.

The tails extending from the two `heads' merge into a a single tail, with a different angle relative to the heads, at a nexus
that corresponds to what we call G229. The high-resolution WISE images show that the extended tail
beyond G229 actually has multiple striations with different position angle. The range of angles may relate 
to an evolution of the relative velocity between the cloud and its local ISM over the past. If the
relative speed is $\sim 3$ \kms\ (discussed below in \S\ref{sec:hydro}), then the evolution in
directions was over the past $10^6$ yr. 

In the 21-cm spectra, there is no clear `wide' spectral line emission that distinguishes itself 
from the rest of the gas, as occurs for the other clouds in this paper. There is a gradual 
trend of velocity, but it is $< 5$ \kms\ and extends over the entire length of the cloud. This trend could
be nothing more than a projection effect due to the inclined velocity vector with respect to the line of sight,
which appears to evolve (from the curved cloud morphology) over the length of the cloud.
Therefore, we can only set an upper limit on the shock velocity. 
If there are shocks, and their `wide' component from the immediate post-shock \hi\ is blended with the 
unshocked gas, then the shock velocity $< 4$ \kms. 

\begin{figure*}
\includegraphics[width=2in]{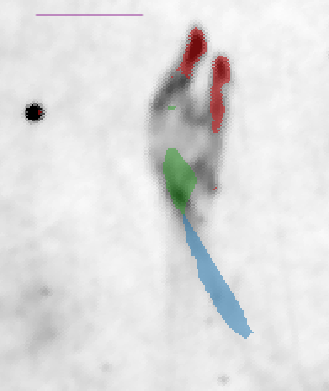}\includegraphics[width=5in]{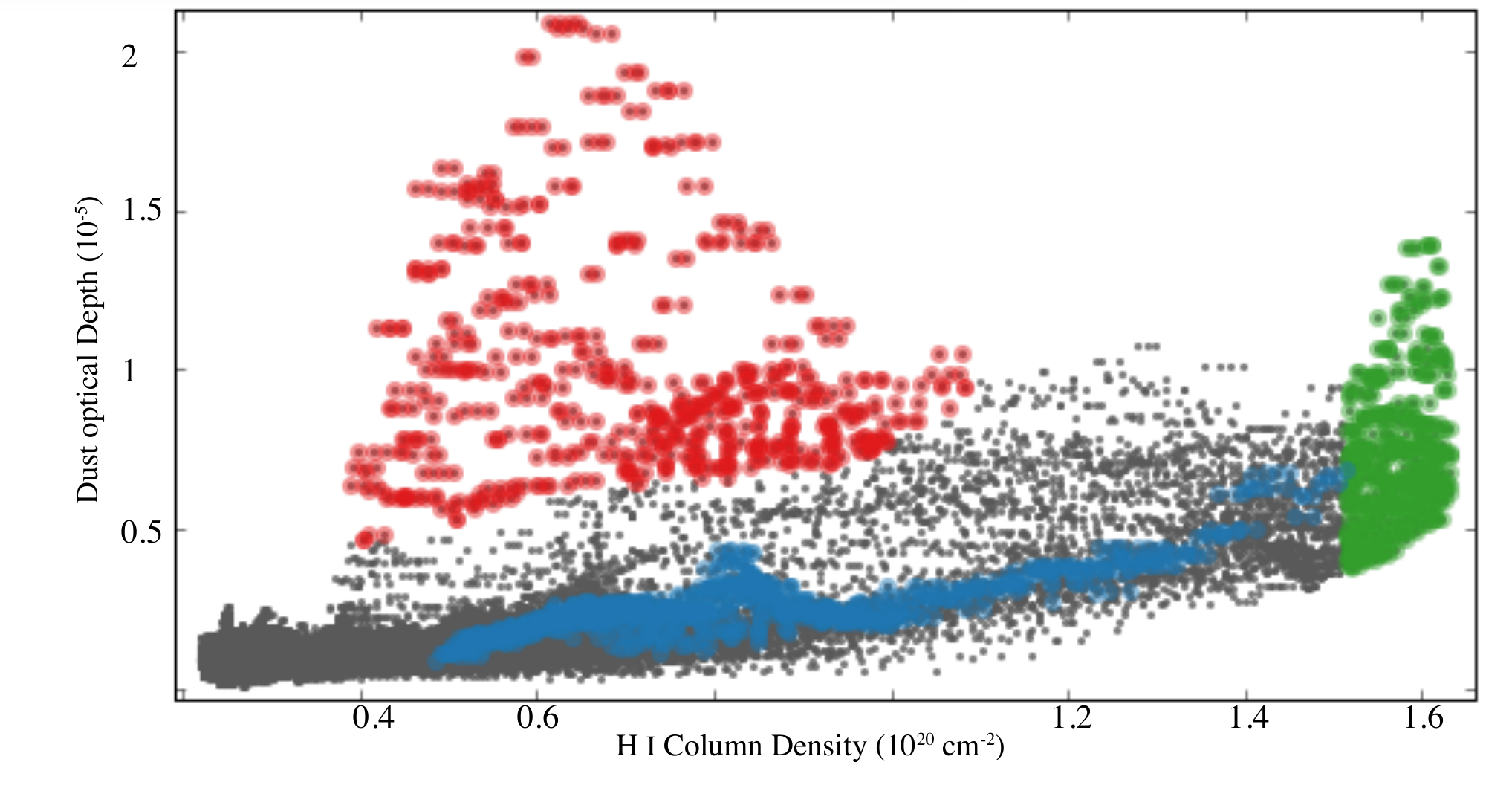}
\caption{Comparison plot of dust optical depth and \hi\ column density for G229 and G225.
The left-hand panel shows the dust image, and the right-hand panel shows the scatter-plot.
The purple scale bar is $1^\circ$ long (1.75 pc if the cloud is at a distance of 100 pc).
Subsets of the data are shown in color. 
In red are the points with the highest excess
of dust optical depth compared to \hi\ column density. Comparing to the left-hand panel, it is evident that all the points with high infrared excess are concentrated into the `heads' of G225 where CO was detected. 
In green are the points selected to have the highest \hi\ column density. Comparing to the left-hand panel, it is evident that the \hi\ peak is in G229, and not coincident with the dust peak.
In blue are the points selected in the southern `tail' of the cloud image south of G229.
Comparing to the right-hand panel, this region has roughly the same dust-versus-gas trend as the rest of the image and is normal, diffuse atomic gas. 
\label{G229and225scat}}
\end{figure*}

\subsection{Atomic and molecular content}

Comparing the dust and gas content of G229 and G225 reveals a very similar trend as seen in
clouds G228 and G230 in the previous section. Figure~\ref{G228and230scat} shows that the cloud
`heads' (including G225) have relatively high dust content. The nexus of the their tails (G229) has the highest 
atomic column density but not the highest overall density as revealed by the {\it Planck} far-infrared data.
The dust emissivity, or ratio of dust optical depth (at 353 GHz) to \hi\ 21-cm column density, is 
$\sigma_{353}=6.7\times 10^{-25}$  cm$^2$ H$^{-1}$ for G225 and $1.0\times 10^{-26}$ cm$^2$ H$^{-1}$ for G229. 
This makes the `tail' emission similar to normal ISM, while the `heads' have much higher dust content
per unit atomic gas. The inferred atomic gas to dust ratios are [G/D]=1.4 in the `head' and
97 in the `tail'. Thus the heads are almost entirely molecular (98\%) while the tails are atomic.

CO observations support this view.
Ground-based observations revealed bright emission with line integrals up to 3 K~km~s$^{-1}$ toward the cloud `heads',
but no detection from the `tails' to an upper limit of $<0.2$ K~km~s$^{-1}$.
If the  total column density (inferred from far-infrared optical depth) of the head is molecular, then $N({\rm H}_2)=8\times 10^{20}$ cm$^{-2}$,
and the ratio of H$_2$ column density to CO line integral is $X=2.7\times 10^{20}$ cm$^{-2}$, somewhat higher than
that of giant molecular clouds. The CO emission is more confined than the far-infrared
emission, with the detected CO being confined to about $6'$. Averaging to $16'$ resolution of the \hi\ survey,
we find that the cloud heads comprise approximately 29\% CO-bright molecular gas, 69\% CO-dark molecular
gas, and 2\% atomic gas.

\section{G243 and G240}

\subsection{Morphology and kinematics}

\begin{figure*}
\includegraphics[width=3in]{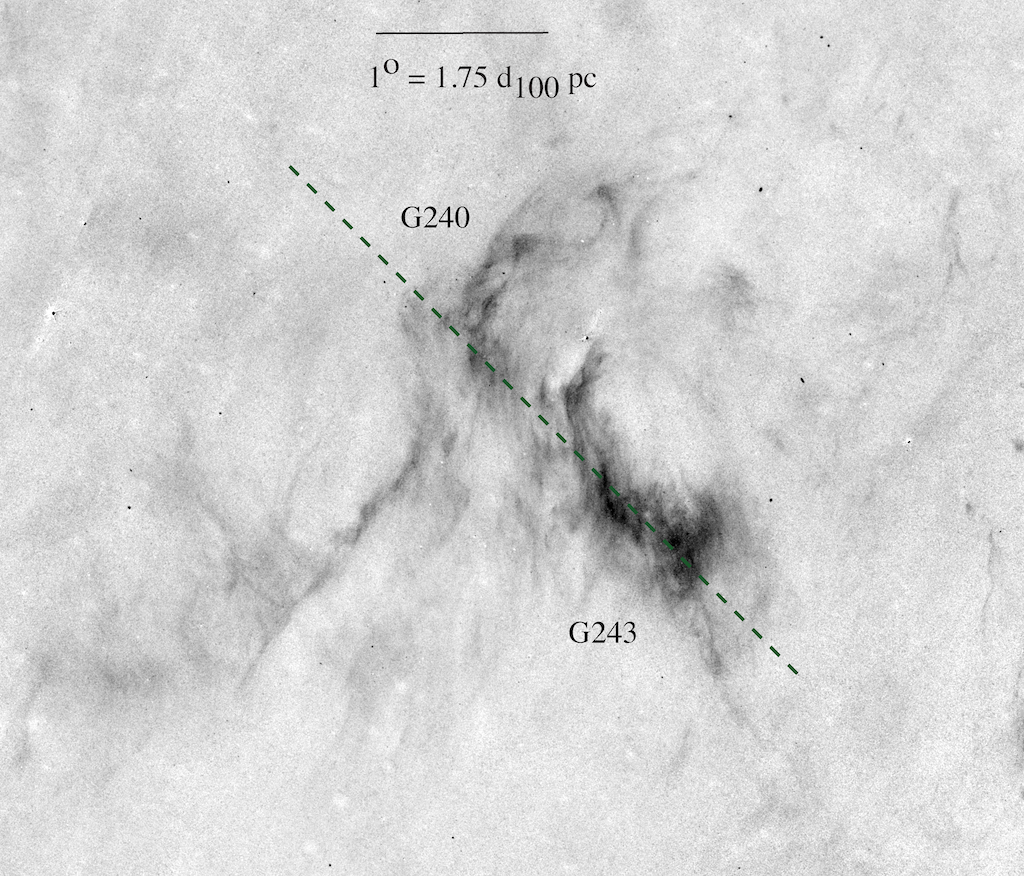}\includegraphics[width=4in]{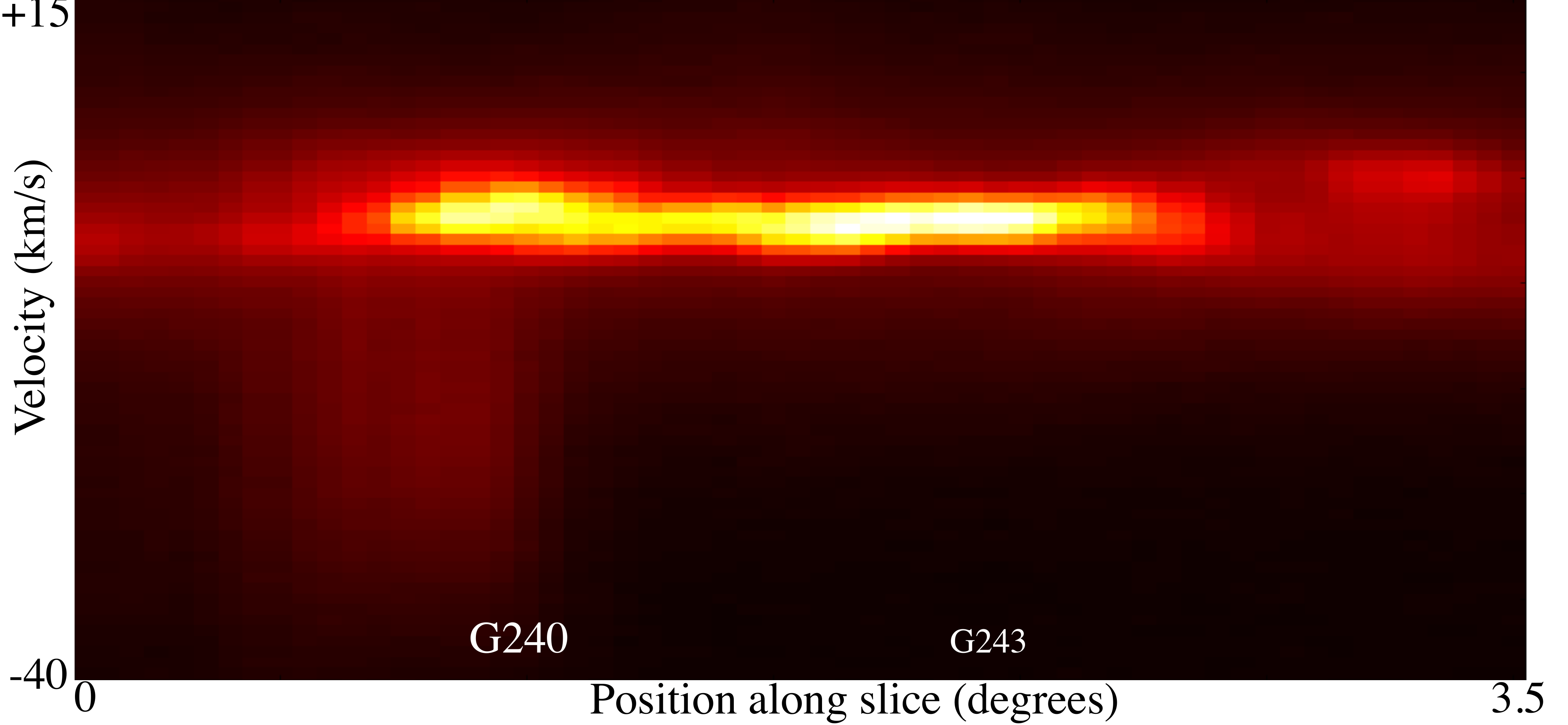}
\caption{Position-velocity slice for clouds G243 and G240. The left-hand panel is the unWISE map of the cloud , with a green, dashed line showing the path along which the \hi\
slice was made. The right-hand panel shows the 21-cm line brightness image, with the horizontal
axis being position along the slice and the vertical axis being the velocity. 
A wide \hi\ component is associated with G240, just upstream of the cloud peak.
\label{G243and240slice}}
\end{figure*}

\begin{figure}
\epsscale{1}
\plotone{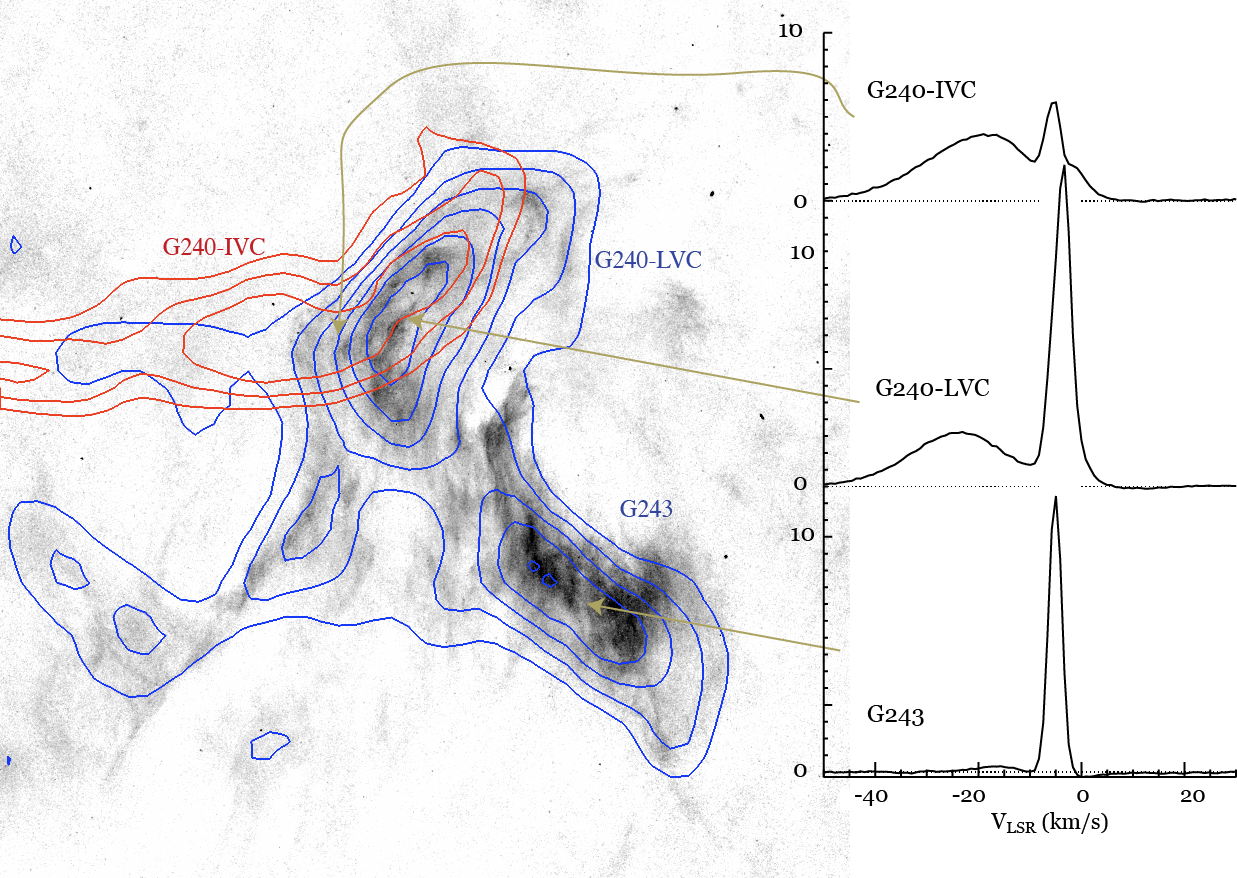}
\caption{
Comparison of the dust (un{\it WISE}) and atomic gas (GASS \hi\ 21-cm line) distributions for the clouds G240 and G243.
The dust map is in greyscale (black meaning more dust), and the atomic gas contours are overlaid. 
The blue contours are integrated over the narrow-component \hi\ velocities (-7 to -3 \kms), showing the bodies of both clouds (G243 and G240-LVC) that dominate the column density. Blue contour levels are linear from 0.5 to $1.6\times 10^{20}$ cm$^{-2}$.
The red contours are the 21-cm line integrated over the intermediate-velocity spectral line (-35 to -9 \kms). Red contour 
levels are linear from 0.5 to $1.5\times 10^{20}$ cm$^{-2}$.
The intermediate-velocity \hi\ is located  upstream of the head of G240 and is labeled G240-IVC.
Note how the shape of the dust cloud filaments are parallel to the interface between the intermediate and low velocity gas in G240.
The right-hand panel shows \hi\ 21-cm spectra for three positions. The top spectrum is toward a position just NE of G240 (2:42:10.5 -35:44:14) where the wide component is relatively prominent. The middle spectrum is centered on the cloud G240
(2:39:23.3 -35:25:02). The lower spectrum is
centered on the cloud G243 (2:34:12.6 -36:56:14). All spectra are averaged 1 beam ($15'$) in the GASS spectral cubes.
\label{G240widenarr}}
\end{figure}

The clouds G243 and G240 were known as G243.2-66.1 and G240.2-65.5, respectively, in HRK. The quality of
new observations from WISE and GASS is higher than was available in 1988, allowing a new assessment and 
potential to identify cloud shocks.
Figure~\ref{G243and240slice} shows the filamentary morphology in the WISE image, with the brightest parts of the cloud comprising a set of curved, approximately parallel filaments. The appearance suggests a set of shocks due to
the cloud moving toward the NE relative to the rest of the local ISM, or equivalently, a shock wave (e.g. from
an old supernova) moving into the cloud from the NE.
The  mass of the complex comprising the two clouds is $58  d_{100}^2  M_\odot$, with
about 11 and 14 $d_{100}^2 M_\odot$ in the concentrations of G240 and  G243, respectively.
The density in the concentrations is approximately 500 $d_{100}^{-1}$ \cm3. 

The \hi\ 21-cm spectra in the G243 and G240 region show a narrow spectral component, with FWHM 3.5 \kms, that
pervades the brightest parts of the filamentary cloud, as well as a broader component.
Figure~\ref{G240widenarr} shows the locations of the broad and narrow emission line components and their spectra.
To remove emission on angular scales larger than the cloud, as well as any residual stray light from far sidelobe 
response, a background spectrum, derived from the lower left corner of the region shown, was removed from all spectra. 
The dust clouds comprise a set of filaments,
with orientation roughly parallel to the interface between the intermediate and low-velocity \hi\ gas.
The wide component is located just NE of (and overlapping with) G240. The configuration is very similar to that seen for G228. 
After subtracting a background spectrum, a fit to the \hi\ line for G240-IVC has FWHM 25 \kms\ and centroid -22 \kms. 
Comparing to the theoretical expectations from \S\ref{sec:theory}, the width of the wide component sets the
 shock velocity $V_s =25$ \kms. 
From Figure~\ref{fig:hih2shock}, the difference between velocities of the IVC and LVC sets $\frac{3}{4}V_s \sin i=17$ \kms,
so the shock is inclined from the line of sight by $25^\circ$ (i.e. close to an edge-on view) and is impacting the cloud on its far side (with the shock moving toward us).

\begin{figure*}
\includegraphics[width=2in]{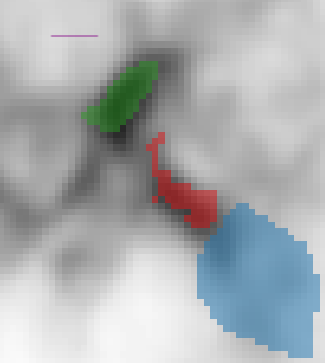}\includegraphics[width=5in]{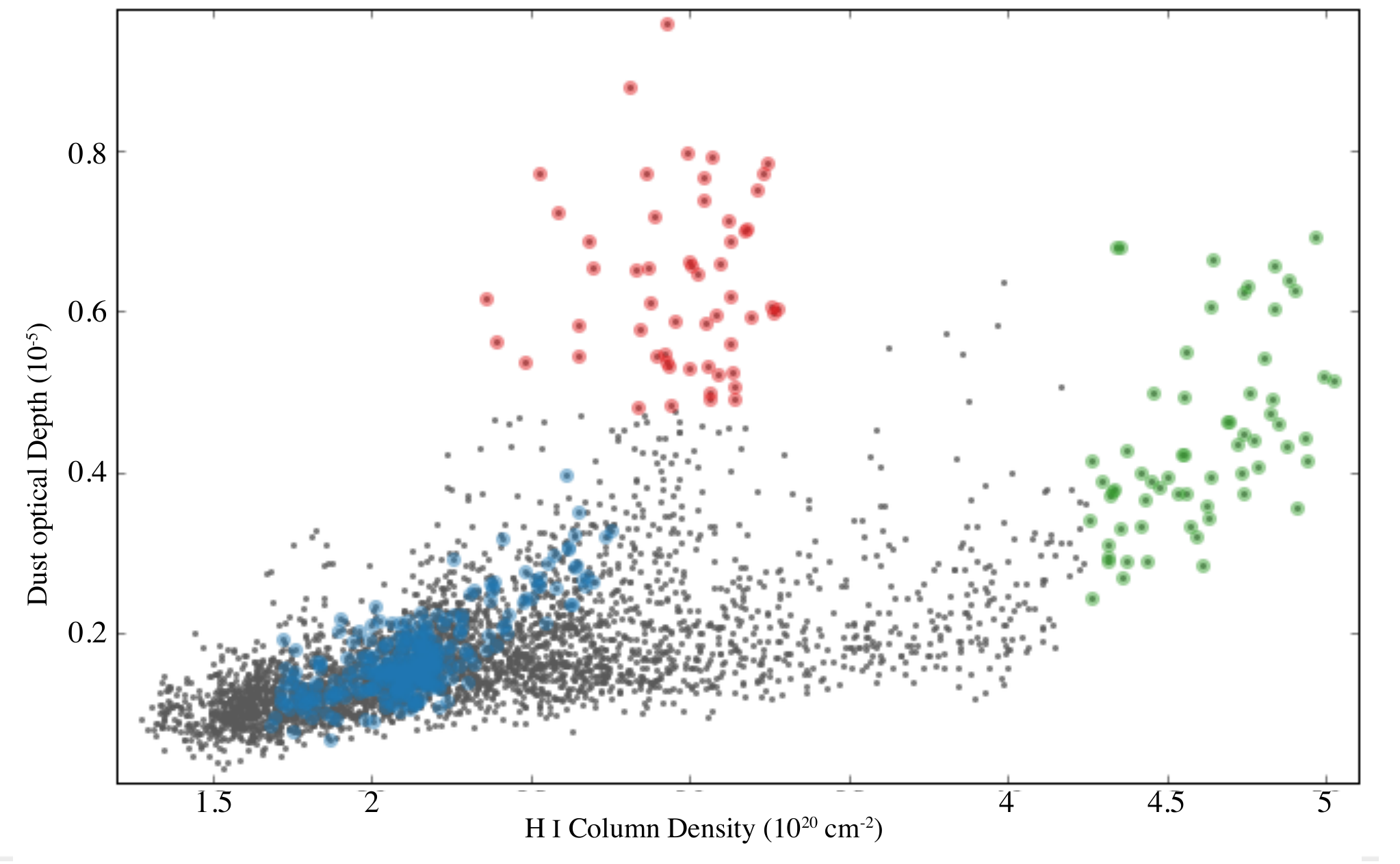}
\caption{Comparison plot of dust optical depth and \hi\ column density for G243 and G240.
The left-hand panel shows the dust image, and the right-hand panel shows the scatter-plot.
The purple scale bar is $1^\circ$ long (1.75 pc if the cloud is at a distance of 100 pc).
Subsets of the data are shown in color. 
In red are the points with the highest excess
of dust optical depth compared to \hi\ column density. Comparing to the left-hand panel, it is evident that
the points with high infrared excess are concentrated in G243.
In green are the points selected to have the highest \hi\ column density. Comparing to the left-hand panel, it is evident that the \hi\ peak is in G240, and not coincident with the dust peak.
In blue are the points selected in the southern `tail' of the cloud image south of G243.
Comparing to the right-hand panel, this region has roughly the same dust-versus-gas trend as the rest of the image and is normal, diffuse atomic gas.
\label{G243and240scat}}
\end{figure*}

\subsection{Atomic and molecular content}

To assess the dust and gas relationship in these clouds, Figure~\ref{G243and240scat} shows the correlation
between the Planck and \hi\ 21-cm column density. 
In G243, the dust emission is significantly in excess relative to the \hi,
and as for the other clouds, the infrared excess is likely tracing H$_2$. 
The G240 portion of the cloud contains the wide \hi\ component. 
Comparing to the theoretical expectations for shocks, we 
interpret G240 as having currently-active shocks fronts, while G243 is all post-shock gas where
H$_2$ has formed. The cooling region is  predicted to be $< 0.1$ pc, much less than the separation
between the clouds, so we do not identify G240 and G243
as portions of the same shock front. The time to cross these clouds at the inferred shock velocity is
 30,000 $d_{100}$ yr, which is comparable to the cloud crossing time.
 This suggests the cloud shocks are, in astronomical terms, relatively recent.

The dust-gas correlation in Figure~\ref{G243and240scat} show that G240, where the present, strong shocks
are likely to be occurring, also has the highest \hi\ column density and dust emission, with approximately
`normal' emissivity based on the grain temperature of 18.5 K \citep{reach17b}, suggesting the G240 is
largely atomic.
The concentration G243 has lower \hi\ column density but higher dust optical depth, leading to an enhanced
emissivity and an inferred H$_2$ column density greater than \hi. There is no CO emission detected 
from these clouds from our ground-based observations (though limited observations were made)
or in the Planck CO map. This suggests that G243 is 72\% CO-dark molecular
gas, and 28\% atomic gas.

\clearpage

\section{Discussion}

\subsection{Summary of Cloud Properties}

\def\tnm{\tablenotemark}
\begin{deluxetable*}{lcccccccc}
\tabletypesize{\scriptsize} 
\tablewidth{0pt}
%\tablenum{text}
\tablecolumns{8}
\tablecaption{Comparative properties of isolated diffuse clouds\label{sumtab}} 
\tablehead{
\colhead{}  &\colhead{}  & \multicolumn{3}{c}{Fraction of mass} & \colhead{$V_s$}  & \colhead{$n({\rm H}) d_{100}$} & &\colhead{$L_{\rm cool}$}\\ \cline{3-5}
\colhead{Cloud }  & \colhead{\hi\ profile} & \colhead{Atomic} & \colhead{Dark} & \colhead{CO-bright}  
& \colhead{(\kms)}  & \colhead{(cm$^{-3}$)} & \colhead{$n/n_0$} & \colhead{(pc)}
}
\startdata
G228.0-28.6   & wide+narrow & 52\% & 26\% & 22\%        & 11  & 600  & 7 & 0.02 \\
G230.1-28.4   & narrow            & 10\% & 80\% & 10\%  &      & 490  &   & 0.02\\
\\
%G236N & 2 narr    & No   & No    & filaments \\
%G236S & narr      & Yes  & YES   & core $7^\prime$\\
%\\
G225.6-66.4  & narrow       &  98\% & 2\% & 0       & $\sim 3$  &5000 & 3 & 0.003 \\%top
G229.0-66.1    & narrow      & 2\% & 69\% &  29\%    &          & 350  &   & 0.009\\%bottom
\\
G240.2-65.6    & wide+narrow    & 90\%  & 10\% & 0    & 25   & 150  & 15 & 0.03 \\
G243.2-66.1     & narrow        & 28\% & 72\% & 0     &      &700  &    & 0.03 \\
\enddata
\end{deluxetable*}

\def\extra{
\begin{deluxetable*}{lccccccccc}
\tabletypesize{\scriptsize} 
\tablewidth{0pt}
%\tablenum{text}
\tablecolumns{9}
\tablecaption{Comparative properties of isolated diffuse clouds\label{sumtab}} 
\tablehead{
\colhead{}  &\colhead{}  & \multicolumn{3}{c}{Fraction of mass} & \colhead{$V_s$} & \colhead{$N_{\rm H}$}  & \colhead{$n({\rm H}) d_{100}$} & &\colhead{$L_{\rm cool}$}\\ \cline{3-5}
\colhead{Cloud }  & \colhead{\hi\ profile} & \colhead{Atomic:Dark:} & \colhead{Dark} & \colhead{CO-bright}  
& \colhead{(\kms)} & \colhead{(cm$^{-2}$)} & \colhead{(cm$^{-3}$)} & \colhead{$n/n_0$} & \colhead{(pc)}
}
\startdata
G228.0-28.6   & wide+narrow & 52\% & 26\% & 22\%        & 11  & $2.1\times 10^{21}$ & 600  & 7 & 0.02 \\
G230.1-28.4   & narrow            & 10\% & 80\% & 10\%  &     & $1.5\times 10^{21}$ & 490  &   & 0.02\\
\\
%G236N & 2 narr    & No   & No    & filaments \\
%G236S & narr      & Yes  & YES   & core $7^\prime$\\
%\\
G225.6-66.4  & narrow       &  98\% & 2\% & 0       & $\sim 3$ & $1.8\times 10^{21}$ &5000 & 3 & 0.003 \\%top
G229.0-66.1    & narrow      & 2\% & 69\% &  29\%    &         & $1.1\times 10^{21}$ & 350  &   & 0.009\\%bottom
\\
G240.2-65.6    & wide+narrow    & 90\%  & 10\% & 0    & 25  & $6.\times 10^{20}$ & 150  & 15 & 0.03 \\
G243.2-66.1     & narrow        & 28\% & 72\% & 0     &     & $8.1\times 10^{20}$ &700  &    & 0.03 \\
\enddata
\end{deluxetable*}
}

Some of the salient properties of the cloud pairs in this study are summarized in Table~\ref{sumtab}. 
We can readily identify some trends in the images and the numerical comparisons of the cloud pairs. 
First, each cloud pair tends to have one component with a filamentary morphology and a wide \hi\ 21-cm
line, while the other component has a more-compact morphology and significant molecular gas.
The wide \hi\ components for G228 and G240  appear just upstream of the clouds 
(Fig.~\ref{G228and230gass} and ~\ref{G229and225slice}) and likely represent the locations of shock fronts.
For those clouds, we use the width of the `wide' component to estimate the shock velocity.
For G225 and G229, no wide component was evident, meaning an upper limit of $\sim 4$ \kms\, consistent
with the hydrodynamic models described below.
The present, average density $n({\rm H})$ was estimated from the the dust column density 
$N({\rm H})+2N({\rm H}_2)$  (assuming gas/dust mass ratio 100) and a cloud depth along the line of sight the same as the geometric mean of 
its length and width in the images.
%; $d_{100}$ is again the distance to the cloud in units of 100 pc. 

The dynamics of the gas are determined by the balance between the specific energies of
gravity (pulling the gas toward its center of mass), 
\begin{equation}
\mathcal{E}_G = \frac{2\pi}{3} G N_{\rm tot} R = 1.0 \times 10^9 \left(\frac{N_{\rm tot}}{10^{21} {\rm cm}^{-2}}\right)
\left(\frac{R}{\rm pc}\right) \, {\rm erg~g}^{-1}
\end{equation}
the flow of the surrounding medium (pushing gas along the direction of the flow), 
\begin{equation}
\mathcal{E}_T = \frac{1}{2} v^2 = 5\times 10^9 \left(\frac{V_s}{{\rm km~s}^{-1}}\right)^2 \, {\rm erg~g}^{-1}
\end{equation}
and magnetic pressure (preventing motion perpendicular to field lines),
\begin{equation}
\mathcal{E}_B = \frac{B^2}{8 \pi \mu m_{\rm H} n} =  1.8\times 10^{10} b^2  \, {\rm erg~g}^{-1}.
\end{equation}
For the `heads' of the cloud pairs in this study, the total proton column density $N_{\rm tot}\sim 10^{21}$ cm$^{-2}$, the size $R\sim d_{100}$ pc, velocity $v\sim 3$--25 km~s$^{-1}$, and we assume $b=1$ $\mu$G cm$^{1.5}$ for the magnetic field scaling. This leads to a proportion 
\begin{equation}
     \mathcal{E}_G : \mathcal {E}_T : \mathcal{E}_B = d_{100} : 4 {V_s}^2 : 12 b^2,
\label{eq:partition}
\end{equation}
where $V_s$ is in \kms, which means that the gas flow through the intercloud medium is the dominant factor, followed by magnetic pressure.
The low proportion for gravity quantifies that these clouds are `diffuse' in the sense of not self-gravitating.
Looking at smaller regions, such as the dense core of G225, where the volume density is much higher than the cloud average, the size is small, so self-gravity remains a minor influence. Magnetic energy is significant and becomes comparable to kinetic energy  for the slow shocks in G225 and G229.

The specific energies $\mathcal{E}_{G,T,B}$ can be directly compared to those derived
from the Millennium Survey of 21-cm absorption and emission for random lines of
sight through the diffuse ISM \citep[][hereafter, HT05]{heilestroland05}, where the energy densities
$E_{T,B}$ were derived. The quantities are simply related by $E=\mathcal{E}/\rho$ where
$\rho$ is the mass density, so the proportions have the same interpretation.
The kinetic specific energy using the turbulent velocity $v_{\rm turb,1D}=1.2$ \kms\ from HT05 is $\mathcal{E}_T=\frac{3}{2}3 v_{turb,1D}^2=2\times 10^{10}$ erg~g$^{-1}$,
which is smaller than we estimate here for the three cloud pairs, because we are estimating 
the {\it flow speed} through the intercloud medium, which drives shocks into
the cloud, as opposed to the the thermal or random motions on small scales
within the clouds. For the magnetic specific energy, HT05 used a fixed
$B=6 \mu$G, which corresponds to $\mathcal{E}_B=6\times 10^{11} n^{-1}$ erg~g$^{-1}$. 
Thus using HT05, 
\begin{equation}
\frac{\mathcal{E}_T}{\mathcal{E}_B} \simeq \left(\frac{n}{30\,{\rm cm}^{-3}}\right),
\end{equation}
which ranges from 5 to 170 for the densities we estimated using the column density divided by cloud size.
On the other hand, from equation~\ref{eq:partition}, we find the ratio
\begin{equation}
\frac{\mathcal{E}_T}{\mathcal{E}_B} = \frac{1}{3} \left(\frac{V_s}{b}\right)^2 ,
\end{equation}
which ranges from 3 to 200 for the flow velocities estimated from the \ion{H}{1} 21-cm
line widths, assuming $b=1$. The local energy densities (or specific energies) are higher using our estimates
than those in HT05, because the present project is focused on individual clouds, while the 
HT05 survey was for random lines of sight that tend to miss the denser regions. 
Also the physical interpretation of the kinetic energy is somewhat different: 
it was considered `turbulent' in HT05 while it is considered a shocked flow
in this work. 
Nonetheless, the {\it ratio} of kinetic to magnetic energy densities is similar
for both approaches, with ratio of turbulent to magnetic energy density (or equivalently, to within
a factor of 2, the plasma $\beta$) greater than unity.
This indicates that for isolated, well-defined clouds, which have
 higher mass density than average, 
gas collisions have a greater influence on gas dynamics than does the magnetic field,
while for the lower-density regions sampled by absorption line surveys, magnetic field dominates.

The properties of the shocks due to the flow through the intercloud medium are elucidated by comparing the 
observed properties of the clouds to
the predictions in \S\ref{sec:theory}. 
For each cloud, we estimate a compression factor from the shock velocity and Table~\ref{modtab}, then divide the present density by that
compression factor to infer the pre-shock velocity. Model shocks were calculated for each cloud's combination of pre-shock density and shock velocity.
In Table~\ref{sumtab}, the length $L_{\rm cool}$ gives the model widths of the shock fronts.
At 100 pc distance, the shock fronts are predicted to subtend approximately $1'$ for G228, G230, G240, and G243 (and even smaller for G225 and G229),
so they are not resolved by the \ion{H}{1} or far-infrared observations presented in this paper.
We discussed above how the thin filaments at the leading edge of G228 seen in the WISE mid-infrared images may be individual shock fronts, 
which may be marginally resolved.

The clouds as a whole are 
much larger than the individual shock fronts for CNM gas. In Figures~\ref{G228and230color}, \ref{G229and225slice}, and \ref{G243and240slice}, 
multiple filamentary structures are evident, so the clouds may comprise a tangled nest of shock fronts. 
The larger-scale and smoother emission
may be from shocks into the lower-density portions of the same cloud. There is no reason to expect the clouds to be
monolithic structures with a uniform density. 
Instead, the lowest-density portions of the clouds will have larger scale lengths.
We assume the shocks are driven by motion of the cloud through the surrounding medium, so the same
shock velocities (3--25 \kms) derived for the shocks into the denser gas, would also apply to the 
lower-density gas. This situation is different from a supernova blast wave \citep[cf.][]{reach19}, where the same
{\it pressure} is impacting clouds.
A shock into a more tenuous part of the cloud
with density 5 cm$^{-3}$ would have a heated layer with thickness of order 1 pc (per eq.~\ref{eq:lcool}), which is the size of the entire cloud and its extended tail. This is consistent with the dynamics being dominated by gas collisions (shocks) for
the main body of the cloud, while the tails would be magnetic dominated (plasma $\beta$ less than 1).

\subsection{Hydrodynamics and Magnetic Field\label{sec:hydro}}

Two-dimensional simulation by \citet{aluzas14} are remarkably similar to the clouds in our sample.
The flow of gas was modeled for cloud pairs with varying configurations and orientation of magnetic field relative to flow velocity.
The initial conditions had two clouds of radius $R$ that were {\it separated}, in the distance perpendicular to the shock velocity, by 0--4 $R$, and {\it offset}, in the direction of the shock velocity, by 0--8 $R$.
The magnetic field orientation was simulated in the directions parallel, perpendicular, and skewed with
respect to the shock velocity. For one-dimensional models, such as discussed in \S\ref{sec:theory},
the component of the magnetic field parallel
to the shock velocity is usually ignored, as it has no effect on the relative ion-neutral dynamics. However, in two dimensions, the effect of the parallel component of the magnetic field becomes evident.

In the  simulations of {\it offset} cloud pairs, the downstream clouds have
significantly different morphology, due to lateral confinement of the gas that reaches the downstream cloud.
A magnetic `flux rope'  forms
behind each cloud, where the magnetic field is focused.
In the simulations of \textit{separated} clouds, each cloud has an individual tail, 
approximately parallel to the other's.  
A particularly interesting situation, relevant to the clouds in
our sample, is  for shock velocity parallel to magnetic field and for cloud
separation $2R$: the tails from the clouds merge
where the plasma $\beta$ is low in the converging flow \citep[see][for the explanation]{maclow94}.

\begin{figure*}
    \centering
    \includegraphics{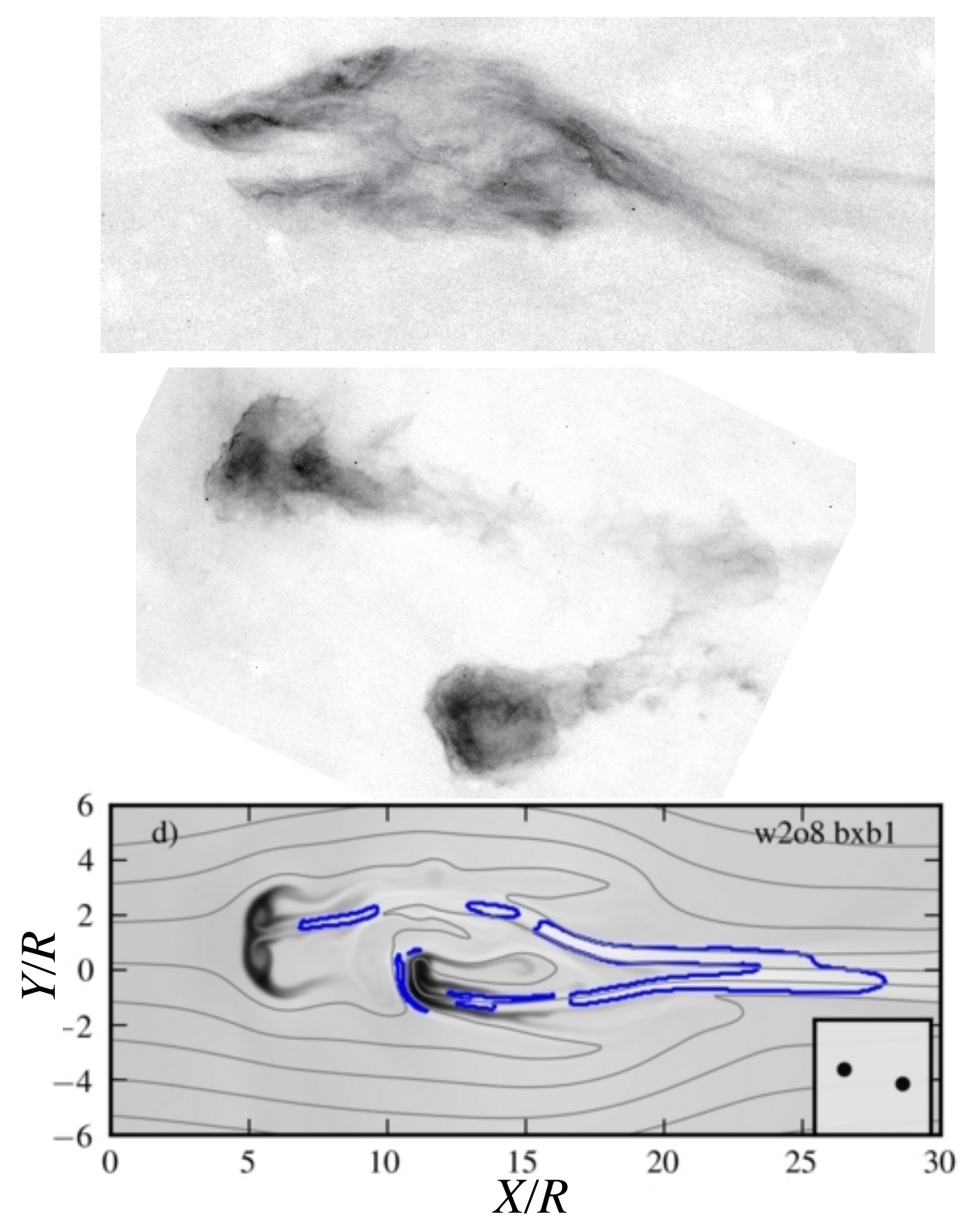}
    \caption{Comparison of MHD simulations \citet[\textit{bottom panel}, from Fig. 2d of][]{aluzas14} with images of two cloud pairs from our sample G228+230 (\textit{middle panel}) and G225+G229 (\textit{top panel}). 
    The cloud images form our sample are the unWISE images, rotated (and inverted for G225+G229) to approximately match the orientation of the simulation. In the simulation, the flow moves from left to right, relative to clouds whose initial configuration is depicted in a small inset. The greyscale is the density and the contours show the magnetic field lines. Blue contours show the flux ropes where plasma $\beta<1$ 
    \citep[for details, see][]{aluzas14}.}
    \label{fig:aluzas}
\end{figure*}

Figure~\ref{fig:aluzas} shows the simulation by \citet{aluzas14} for separation $2R$ and offset $8R$,
with magnetic field parallel to shock velocity.
Note the merged tails, which are similar to those observed in the cloud pairs G225+G229 as well 
as G228+230. 
For G225+229, based upon the lack of \ion{H}{1} accelerated to more than  5 \kms\ and
the remarkable similarity of the MHD simulation with the observations, we assign the 
shock properties based upon the MHD simulation input, which was a sonic Mach number of 3. 
If the clouds had CNM temperatures $\sim 100$ K, then the shock velocity is $\sim 3$ \kms.
Inspecting the unWISE and {\it Planck} images, the tails are not straight, instead having a
significant twist. This effect is also seen in simulations where the magnetic field is skewed
with respect to the shock velocity \citep[Fig.~5 of][]{aluzas14}. 

The cloud pair G243+G240 has some similarity to the MHD simulation  with
zero lateral separation but $8R$ offset between the clouds \citep[Fig.~2e of][]{aluzas14}.
In this case the magnetic flux rope from the upstream cloud falls onto
the downstream cloud, which is more compressed. 
The large \hi\ line widths indicate shock velocities much higher than the MHD simulations, but the
general principle of the downstream cloud being more compressed may still apply.
This extra compression may contribute to making 
the downstream cloud (G243) molecular, as evidenced by its elevated dust column density
per unit \ion{H}{1} column density (red points in Figure~\ref{G243and240scat} of this paper).

The MHD simulations explain what had been a puzzling morphology for high-latitude clouds.
Whereas one-dimensional thinking  led us to expect the downstream gas from each cloud would evolve  independently, in fact the parallel component of the magnetic field has a significant effect on the flow,
leading to merged tails  behind cloud pairs. The simulations also give some insight into the potential age of
of the flow. Figure~3 from \citet{aluzas14} shows the $2R$ separation and $8R$ offset case at
different times. The merged-tail (joined flux rope) morphology occurs in the age range 2--5 times the
cloud crushing time, defined as 10 $R/v_s$ for the simulated clouds which had density contrast 100 relative
to the intercloud gas. Using the radius and shock velocity of the clouds in our sample, 
the age range for the merged tails is 0.4--5 Myr. This age is much longer than the cooling time
for individual shocks. The scenario for the clouds can then be understood as a long-term interaction with
intercloud gas leading to their overall morphology.

\begin{figure}
    \centering
    \includegraphics{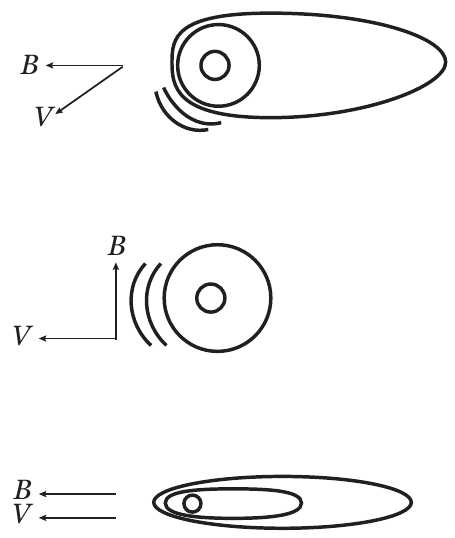}
    \caption{Illustration of cloud shape depending upon the orientation of the motion of the cloud through the intercloud medium
    (arrows labeled {\it V}) and the magnetic field (arrows labeled {\it B}). The arcs show the locations of non-dissociative
    C-shocks, and the elongated ellipses show the locations of cloud `tails' such as found in the MHD models, due to material
    swept down stream along magnetic field lines.}
    \label{fig:magvel}
\end{figure}

Some of the varying shapes of diffuse molecular clouds are determined by the relative orientations between the flow through
the intercloud medium and the magnetic field. Figure~\ref{fig:magvel} synthesizes what we learned from comparing the observations to
the MHD models.
The high-compression C-type shocks form where the magnetic field is perpendicular to the shock \S\ref{sec:theory}, 
while the long tails are drawn out parallel to the magnetic field.

{\bfc 
In this discussion, we compared the observed properties of the clouds, from two-dimensional (2D) images, with MHD simulations that were also 2D. The 2D simulations lack vorticity transport that
occurs in 3D reality;  the effect of magnetic fields may be artificially enhanced and instabilities 
suppressed. Other simulations on larger scale show dynamic formation of clouds in the range
of sizes of those considered in this study \citep{audit05,heitsch11}, though those studies are
also 2D and do not include magnetic fields. Future simulations extending to 3D, with comparison to
the observed properties of clouds, are likely to advance our understanding to the longevity
of clouds and their small-scale structure.
}
%\clearpage

\subsection{Implications for Formation of Diffuse Molecular Clouds}

The observations of cloud geometry and kinematics, in comparison to the MHD simulations, indicate clouds with field parallel to shock velocity. Why do we tend to see this orientation?
The answer may lie as much in the {\it formation} of the clouds as in their evolution.
\citet{hennebelle00} showed that cold (CNM) clouds condense out of a warm medium (WNM)
under conditions that lead to alignment between the magnetic field and the flow. 
For the `transverse' case where field is perpendicular to flow, the magnetic field becomes compressed and resists condensation, so that a CNM cloud does not form.
In contrast, for the case where field is more nearly parallel to flow, the magnetic tension generates 
transverse velocities that `unbend' the field, leading to condensation of a CNM cloud.
The field and flow alignment need not be precise; condensation 
occurs  for alignment within $20^\circ$.

\citet{hennebelle00} further divide the parallel case into 
a weak field case, where the magnetic field becomes aligned with the flow, and
a strong field case, where the flow becomes aligned with the field; in either of these cases, the result is toward
alignment between field and flow.
The weak field case applies when
\begin{equation}
    B^2 < \mu m_{\rm H} n_w v^2
\end{equation}
where $n_w$ is the WNM density, $v$ is the colliding flow speed, and $B_x$ is the magnetic field component strength. If we use the empirically observed scaling of magnetic field strength with density for
interstellar clouds (extrapolating from larger clouds to the 100 $M_\odot$ clouds considered in this study), 
$B = b n^\frac{1}{2}$, 
with $b\simeq 1$ for $B$ in $\mu$G and $n$ in cm$^{-3}$, then
the weak field condition becomes only a function of the flow speed
\begin{equation}
    v > 7 b^{0.5} {\rm ~km~s}^{-1}.
\end{equation}
For the  shock properties summarized in Table~\ref{sumtab}, the strong field case to apples to
the G225+G229 cloud pair, which may explain why Figure~\ref{fig:aluzas} shows such a strong resemblance 
to the 
parallel, magnetic field dominated models by \citet{aluzas14}. 
The G228+G230 cloud pair is in the weak-strong transition.
The G240+G243 cloud pair is in the weak-field case; indeed the Mach number is much higher for these clouds than the
theoretical calculations.

We showed that the CNM clouds have evidence of shocks and of a high fraction of H$_2$,
both for the cloud pairs in this paper and a more systematically-selected set of clouds \citep{reach17}.
Numerical simulations by \citet{clark19} showed that 
Mach $\simgt$ 2 collisions of $n_0=10$ cm$^{-3}$ atomic clouds result in efficient conversion of 
initially atomic gas to 50\% H$_2$.
The initial conditions used by \citet{clark19} of two identical clouds colliding head-on are not
the same as what we envision from the appearance of the shocked cloud pairs presented in this paper,
where the clouds appear more dense and are interacting with (or condensed from) a lower-density
intercloud medium. Nonetheless the physical properties of the shocked clouds are relevant, and
the results are in general accord with the observations  presented in this paper.
The shocked clouds appear to have a large fraction of H$_2$ that is `CO dark' (cf. Figs.~\ref{G228and230scat}, \ref{G229and225scat}, and \ref{G243and240scat}).
\def\extra{\citet{clark19} calculate the observable emission
signatures of the state of C in the shocked clouds, and they showed that
[\ion{C}{2}] far-infrared emission is the dominant tracer up to gas densities of $10^3$ cm$^{-3}$,  while [\ion{C}{1}] and CO become important only in gas denser than $10^4$ cm$^{-3}$.}

%\clearpage
\section{Conclusions}

Shocks into diffuse clouds should be common, based on their wide range of velocities through
their local intercloud medium.
Tracers of shocks are observed via absorption lines by transient species sometimes requiring endothermic
reactions to form (like CH$^+$). Absorption lines sample random pencil beams through the ISM and do not
readily map to individual cloud structures. Taking a different approach, we utilized wide area surveys of
\hi\ 21-cm and dust far-infrared emission to identify pairs of shocked clouds. 
From 1-dimensional shock models, we showed that primary signature of individual shocks from those data would be 
dynamical, from the width of 21-cm lines and their configuration compared to morphology and
inferred shock location and direction.

In all three cloud pairs, there was morphology where a leading
edge of the cloud (i.e. the part directly facing the incoming gas, in the rest frame of the cloud) was
relatively sharply defined edges, with `heads'  of size $\sim 0.5$  pc. 
One potential feature of cloud pairs is that they may be interacting with {\it each other}, although
for our cloud sample this does not appear to be the case. The head/tail morphology is roughly parallel for
both components in each  cloud pair.
The high-resolution WISE images show
intricate, filamentary morphology; each filament is likely either an active or recent-past shock front.
The cloud heads are largely molecular, even where no CO was detected, with 'CO-dark' molecular gas
comprising up to 80\% of the mass.
All three cloud pairs have diffuse `tails' of atomic gas extend for 1-4 pc behind them. 

The shocks in the three cloud pairs in this study range from strong a strong shock with Mach number $\sim 25$ for
G240+G243 to a moderately strong shock with Mach number $\sim 11$ for G228+230 to a weak shock with Mach number $\sim 3$ for G225+G229. The slower shocks are C-type and were modeled with the Paris-Durham shock code, which provides
a good explanation of the \ion{H}{1} 21-cm line shapes.

The cloud pairs we described should be considered `normal' diffuse interstellar clouds, despite the  range of shock velocities. The present shocks, of which we are detecting direct evidence, are likely
not the first shock any given cloud has experienced. This explains why the dust/gas ratio and dust properties
(inferred from temperature of big grains or very small grain abundance) are similar for most of
the diffuse interstellar medium. Essentially, all of the diffuse interstellar medium has experienced
shocks with in the range of velocities identified in this study, and this shock history
is an essential part of understanding the properties of interstellar material.

\acknowledgements  
This publication makes use of data products from the Wide-field Infrared Survey Explorer, which is a joint project of the University of California, Los Angeles, and the Jet Propulsion Laboratory/California Institute of Technology, funded by the National Aeronautics and Space Administration.
This work made use of the open-source python packages {\tt matplotlib} \citep{matplotlib} and {\tt astropy} \citep{astropy}. The Parkes radio telescope is part of the Australia Telescope National Facility which is funded by the Australian Government for operation as a National Facility managed by CSIRO; we acknowledge the Wiradjuri people as the traditional owners of that Observatory site.

\facilities{Planck, WISE, Parkes}
\clearpage

\bibliography{wtrbib}

\begin{thebibliography}{}
\expandafter\ifx\csname natexlab\endcsname\relax\def\natexlab#1{#1}\fi

\bibitem[{{Al{\={u}}zas} {et~al.}(2014){Al{\={u}}zas}, {Pittard}, {Falle}, \&
  {Hartquist}}]{aluzas14}
{Al{\={u}}zas}, R., {Pittard}, J.~M., {Falle}, S.~A.~E.~G., \& {Hartquist},
  T.~W. 2014, \mnras, 444, 971

\bibitem[{{Astropy Collaboration} {et~al.}(2018){Astropy Collaboration},
  {Price-Whelan}, {Sip{\H{o}}cz}, {G{\"u}nther}, {Lim}, {Crawford}, {Conseil},
  {Shupe}, {Craig}, {Dencheva}, {Ginsburg}, {Vand erPlas}, {Bradley},
  {P{\'e}rez-Su{\'a}rez}, {de Val-Borro}, {Aldcroft}, {Cruz}, {Robitaille},
  {Tollerud}, {Ardelean}, {Babej}, {Bach}, {Bachetti}, {Bakanov}, {Bamford},
  {Barentsen}, {Barmby}, {Baumbach}, {Berry}, {Biscani}, {Boquien}, {Bostroem},
  {Bouma}, {Brammer}, {Bray}, {Breytenbach}, {Buddelmeijer}, {Burke},
  {Calderone}, {Cano Rodr{\'\i}guez}, {Cara}, {Cardoso}, {Cheedella}, {Copin},
  {Corrales}, {Crichton}, {D'Avella}, {Deil}, {Depagne}, {Dietrich}, {Donath},
  {Droettboom}, {Earl}, {Erben}, {Fabbro}, {Ferreira}, {Finethy}, {Fox},
  {Garrison}, {Gibbons}, {Goldstein}, {Gommers}, {Greco}, {Greenfield},
  {Groener}, {Grollier}, {Hagen}, {Hirst}, {Homeier}, {Horton}, {Hosseinzadeh},
  {Hu}, {Hunkeler}, {Ivezi{\'c}}, {Jain}, {Jenness}, {Kanarek}, {Kendrew},
  {Kern}, {Kerzendorf}, {Khvalko}, {King}, {Kirkby}, {Kulkarni}, {Kumar},
  {Lee}, {Lenz}, {Littlefair}, {Ma}, {Macleod}, {Mastropietro}, {McCully},
  {Montagnac}, {Morris}, {Mueller}, {Mumford}, {Muna}, {Murphy}, {Nelson},
  {Nguyen}, {Ninan}, {N{\"o}the}, {Ogaz}, {Oh}, {Parejko}, {Parley}, {Pascual},
  {Patil}, {Patil}, {Plunkett}, {Prochaska}, {Rastogi}, {Reddy Janga},
  {Sabater}, {Sakurikar}, {Seifert}, {Sherbert}, {Sherwood-Taylor}, {Shih},
  {Sick}, {Silbiger}, {Singanamalla}, {Singer}, {Sladen}, {Sooley},
  {Sornarajah}, {Streicher}, {Teuben}, {Thomas}, {Tremblay}, {Turner},
  {Terr{\'o}n}, {van Kerkwijk}, {de la Vega}, {Watkins}, {Weaver}, {Whitmore},
  {Woillez}, {Zabalza}, \& {Astropy Contributors}}]{astropy}
{Astropy Collaboration}, {Price-Whelan}, A.~M., {Sip{\H{o}}cz}, B.~M., {et~al.}
  2018, \aj, 156, 123

\bibitem[{{Audit} \& {Hennebelle}(2005)}]{audit05}
{Audit}, E., \& {Hennebelle}, P. 2005, \aap, 433, 1

\bibitem[{{Bacalla} {et~al.}(2019){Bacalla}, {Linnartz}, {Cox}, {Cami},
  {Roueff}, {Smoker}, {Farhang}, {Bouwman}, \& {Zhao}}]{bacalla19}
{Bacalla}, X.~L., {Linnartz}, H., {Cox}, N. L.~J., {et~al.} 2019, \aap, 622,
  A31

\bibitem[{{Beaumont} {et~al.}(2015){Beaumont}, {Goodman}, \&
  {Greenfield}}]{beaumont15}
{Beaumont}, C., {Goodman}, A., \& {Greenfield}, P. 2015, in Astronomical
  Society of the Pacific Conference Series, Vol. 495, Astronomical Data
  Analysis Software an Systems XXIV (ADASS XXIV), ed. A.~R. {Taylor} \&
  E.~{Rosolowsky} (Astronomical Society of the Pacific), 101

\bibitem[{{Beichman} {et~al.}(1988){Beichman}, {Neugebauer}, {Habing}, Clegg,
  \& Chester}]{irasexpsupp}
{Beichman}, C.~A., {Neugebauer}, G., {Habing}, H.~J., Clegg, P.~E., \& Chester,
  T.~J. 1988, {Infrared Astronomical Satellite (IRAS) Catalogs and Atlases.
  Volume 1: Explanatory Supplement}, Vol.~1 (NASA: Washington, DC)

\bibitem[{{Binney} \& {Merrifield}(1998)}]{binney98}
{Binney}, J., \& {Merrifield}, M. 1998, Galactic Astronomy (Princeton
  University Press)

\bibitem[{{Bolatto} {et~al.}(2013){Bolatto}, {Wolfire}, \&
  {Leroy}}]{xfactorreview}
{Bolatto}, A.~D., {Wolfire}, M., \& {Leroy}, A.~K. 2013, \araa, 51, 207

\bibitem[{{Chevance} {et~al.}(2020){Chevance}, {Madden}, {Fischer}, {Vacca},
  {Lebouteiller}, {Fadda}, {Galliano}, {Indebetouw}, {Kruijssen}, {Lee},
  {Poglitsch}, {Polles}, {Cormier}, {Hony}, {Iserlohe}, {Krabbe}, {Meixner},
  {Sabbi}, \& {Zinnecker}}]{chevance20}
{Chevance}, M., {Madden}, S.~C., {Fischer}, C., {et~al.} 2020, \mnras, 494,
  5279

\bibitem[{{Clark} {et~al.}(2019){Clark}, {Glover}, {Ragan}, \&
  {Duarte-Cabral}}]{clark19}
{Clark}, P.~C., {Glover}, S. C.~O., {Ragan}, S.~E., \& {Duarte-Cabral}, A.
  2019, \mnras, 486, 4622

\bibitem[{{Cox}(1979)}]{cox79}
{Cox}, D.~P. 1979, \apj, 234, 863

\bibitem[{{Dickey} \& {Lockman}(1990)}]{dickeylockman}
{Dickey}, J.~M., \& {Lockman}, F.~J. 1990, \araa, 28, 215

\bibitem[{{Draine} \& {McKee}(1993)}]{drainemckee93}
{Draine}, B.~T., \& {McKee}, C.~F. 1993, \araa, 31, 373

\bibitem[{{Elitzur} \& {Watson}(1980)}]{elitzur80}
{Elitzur}, M., \& {Watson}, W.~D. 1980, \apj, 236, 172

\bibitem[{{Fahrion} {et~al.}(2017){Fahrion}, {Cormier}, {Bigiel}, {Hony},
  {Abel}, {Cigan}, {Csengeri}, {Graf}, {Lebouteiller}, {Madden}, {Wu}, \&
  {Young}}]{fahrion17}
{Fahrion}, K., {Cormier}, D., {Bigiel}, F., {et~al.} 2017, \aap, 599, A9

\bibitem[{{Falgarone} {et~al.}(2009){Falgarone}, {Pety}, \&
  {Hily-Blant}}]{falgarone09}
{Falgarone}, E., {Pety}, J., \& {Hily-Blant}, P. 2009, \aap, 507, 355

\bibitem[{{Field} {et~al.}(1969){Field}, {Goldsmith}, \& {Habing}}]{field69}
{Field}, G.~B., {Goldsmith}, D.~W., \& {Habing}, H.~J. 1969, \apjl, 155, L149

\bibitem[{Fitzpatrick(2014)}]{fitzpatrick14}
Fitzpatrick, R. 2014, Plasma Physics: An Introduction No. ISBN
  978-1-4665-9426-5 (CRC Press, Taylor \& Francis Group)

\bibitem[{{Flower} \& {Pineau des For{\^e}ts}(2003)}]{flowerpineau03}
{Flower}, D.~R., \& {Pineau des For{\^e}ts}, G. 2003, \mnras, 343, 390

\bibitem[{{Flower} \& {Pineau des For{\^e}ts}(2015)}]{flower15}
---. 2015, \aap, 578, A63

\bibitem[{{Frisch} {et~al.}(1999){Frisch}, {Dorschner}, {Geiss}, {Greenberg},
  {Gr{\"u}n}, {Landgraf}, {Hoppe}, {Jones}, {Kr{\"a}tschmer}, {Linde},
  {Morfill}, {Reach}, {Slavin}, {Svestka}, {Witt}, \& {Zank}}]{frisch99}
{Frisch}, P.~C., {Dorschner}, J.~M., {Geiss}, J., {et~al.} 1999, \apj, 525, 492

\bibitem[{{Godard} {et~al.}(2009){Godard}, {Falgarone}, \& {Pineau Des
  For{\^e}ts}}]{godard09}
{Godard}, B., {Falgarone}, E., \& {Pineau Des For{\^e}ts}, G. 2009, \aap, 495,
  847

\bibitem[{{Green} {et~al.}(2019){Green}, {Schlafly}, {Zucker}, {Speagle}, \&
  {Finkbeiner}}]{green19}
{Green}, G.~M., {Schlafly}, E., {Zucker}, C., {Speagle}, J.~S., \&
  {Finkbeiner}, D. 2019, \apj, 887, 93

\bibitem[{{Grenier} {et~al.}(2005){Grenier}, {Casandjian}, \&
  {Terrier}}]{grenier05}
{Grenier}, I.~A., {Casandjian}, J.-M., \& {Terrier}, R. 2005, Science, 307,
  1292

\bibitem[{{Haisch} {et~al.}(2001){Haisch}, {Lada}, \& {Lada}}]{haisch01}
{Haisch}, Karl~E., J., {Lada}, E.~A., \& {Lada}, C.~J. 2001, \apjl, 553, L153

\bibitem[{Heiles \& Crutcher(2005)}]{heilescrutcher05}
Heiles, C., \& Crutcher, R. 2005, in Cosmic Magnetic Fields, ed. R.~Wielebinski
  \& R.~Beck, Lecture Notes in Physics (Berlin, Heidelberg: Springer Berlin
  Heidelberg), 137--182

\bibitem[{{Heiles} \& {Habing}(1974)}]{heileshabing}
{Heiles}, C., \& {Habing}, H.~J. 1974, \aaps, 14, 1

\bibitem[{{Heiles} {et~al.}(1988){Heiles}, {Reach}, \& {Koo}}]{hrk88}
{Heiles}, C., {Reach}, W.~T., \& {Koo}, B.-C. 1988, ApJ, 332, 313

\bibitem[{{Heiles} \& {Troland}(2003)}]{heilestroland03b}
{Heiles}, C., \& {Troland}, T.~H. 2003, \apj, 586, 1067

\bibitem[{{Heiles} \& {Troland}(2005)}]{heilestroland05}
---. 2005, \apj, 624, 773

\bibitem[{{Heitsch} {et~al.}(2011){Heitsch}, {Naab}, \& {Walch}}]{heitsch11}
{Heitsch}, F., {Naab}, T., \& {Walch}, S. 2011, \mnras, 415, 271

\bibitem[{{Hennebelle} \& {P{\'e}rault}(2000)}]{hennebelle00}
{Hennebelle}, P., \& {P{\'e}rault}, M. 2000, \aap, 359, 1124

\bibitem[{{Hollenbach} \& {McKee}(1989)}]{hm89}
{Hollenbach}, D., \& {McKee}, C.~F. 1989, \apj, 342, 306

\bibitem[{Hunter(2007)}]{matplotlib}
Hunter, J.~D. 2007, Computing in Science \& Engineering, 9, 90

\bibitem[{{Indriolo} \& {McCall}(2012)}]{indriolo12}
{Indriolo}, N., \& {McCall}, B.~J. 2012, \apj, 745, 91

\bibitem[{{Ingalls} {et~al.}(2011){Ingalls}, {Bania}, {Boulanger}, {Draine},
  {Falgarone}, \& {Hily-Blant}}]{ingalls11}
{Ingalls}, J.~G., {Bania}, T.~M., {Boulanger}, F., {et~al.} 2011, \apj, 743,
  174

\bibitem[{{Kalberla} \& {Haud}(2015)}]{kalberla15}
{Kalberla}, P.~M.~W., \& {Haud}, U. 2015, \aap, 578, A78

\bibitem[{{Kalberla} {et~al.}(2020){Kalberla}, {Kerp}, \& {Haud}}]{kalberla20}
{Kalberla}, P.~M.~W., {Kerp}, J., \& {Haud}, U. 2020, arXiv e-prints,
  arXiv:2004.14630

\bibitem[{{Krumholz} {et~al.}(2019){Krumholz}, {McKee}, \& {Bland
  -Hawthorn}}]{krumholz19}
{Krumholz}, M.~R., {McKee}, C.~F., \& {Bland -Hawthorn}, J. 2019, \araa, 57,
  227

\bibitem[{{Kulkarni} \& {Heiles}(1988)}]{kulkarniheiles}
{Kulkarni}, S.~R., \& {Heiles}, C. 1988, in Galactic and Extragalactic Radio
  Astronomy, ed. K.~I. {Kellermann} \& G.~L. {Verschuur}, 95--153

\bibitem[{{Langer} {et~al.}(2014){Langer}, {Velusamy}, {Pineda}, {Willacy}, \&
  {Goldsmith}}]{langer14}
{Langer}, W.~D., {Velusamy}, T., {Pineda}, J.~L., {Willacy}, K., \&
  {Goldsmith}, P.~F. 2014, \aap, 561, A122

\bibitem[{{Lee} {et~al.}(2015){Lee}, {Stanimirovi{\'c}}, {Murray}, {Heiles}, \&
  {Miller}}]{lee15}
{Lee}, M.-Y., {Stanimirovi{\'c}}, S., {Murray}, C.~E., {Heiles}, C., \&
  {Miller}, J. 2015, \apj, 809, 56

\bibitem[{{Leroy} {et~al.}(2007){Leroy}, {Bolatto}, {Stanimirovic}, {Mizuno},
  {Israel}, \& {Bot}}]{leroy07}
{Leroy}, A., {Bolatto}, A., {Stanimirovic}, S., {et~al.} 2007, \apj, 658, 1027

\bibitem[{{Lesaffre} {et~al.}(2013){Lesaffre}, {Pineau des For{\^e}ts},
  {Godard}, {Guillard}, {Boulanger}, \& {Falgarone}}]{lesaffre13}
{Lesaffre}, P., {Pineau des For{\^e}ts}, G., {Godard}, B., {et~al.} 2013, \aap,
  550, A106

\bibitem[{{Li} {et~al.}(2015){Li}, {Ostriker}, {Cen}, {Bryan}, \&
  {Naab}}]{liostriker15}
{Li}, M., {Ostriker}, J.~P., {Cen}, R., {Bryan}, G.~L., \& {Naab}, T. 2015,
  \apj, 814, 4

\bibitem[{{Low} {et~al.}(1984){Low}, {Young}, {Beintema}, {Gautier},
  {Beichman}, {Aumann}, {Gillett}, {Neugebauer}, {Boggess}, \&
  {Emerson}}]{low84}
{Low}, F.~J., {Young}, E., {Beintema}, D.~A., {et~al.} 1984, \apjl, 278, L19

\bibitem[{{Luhman}(2012)}]{luhman12}
{Luhman}, K.~L. 2012, \araa, 50, 65

\bibitem[{{Mac Low} {et~al.}(1994){Mac Low}, {McKee}, {Klein}, {Stone}, \&
  {Norman}}]{maclow94}
{Mac Low}, M.-M., {McKee}, C.~F., {Klein}, R.~I., {Stone}, J.~M., \& {Norman},
  M.~L. 1994, \apj, 433, 757

\bibitem[{{Magnani} \& {Smith}(2010)}]{magnani10}
{Magnani}, L., \& {Smith}, A.~J. 2010, \apj, 722, 1685

\bibitem[{{Mathis} {et~al.}(1983){Mathis}, {Mezger}, \& {Panagia}}]{mmp83}
{Mathis}, J.~S., {Mezger}, P.~G., \& {Panagia}, N. 1983, \aap, 500, 259

\bibitem[{{McClure-Griffiths} {et~al.}(2009){McClure-Griffiths}, {Pisano},
  {Calabretta}, {Ford}, {Lockman}, {Staveley-Smith}, {Kalberla}, {Bailin},
  {Dedes}, {Janowiecki}, {Gibson}, {Murphy}, {Nakanishi}, \&
  {Newton-McGee}}]{mcclure09}
{McClure-Griffiths}, N.~M., {Pisano}, D.~J., {Calabretta}, M.~R., {et~al.}
  2009, \apjs, 181, 398

\bibitem[{{McKee} \& {Ostriker}(1977)}]{mckeeostriker}
{McKee}, C.~F., \& {Ostriker}, J.~P. 1977, \apj, 218, 148

\bibitem[{{Meisner} \& {Finkbeiner}(2014)}]{meisner14}
{Meisner}, A.~M., \& {Finkbeiner}, D.~P. 2014, \apj, 781, 5

\bibitem[{{Murray} {et~al.}(2018{\natexlab{a}}){Murray}, {Peek}, {Lee}, \&
  {Stanimirovi{\'c}}}]{murray18a}
{Murray}, C.~E., {Peek}, J.~E.~G., {Lee}, M.-Y., \& {Stanimirovi{\'c}}, S.
  2018{\natexlab{a}}, \apj, 862, 131

\bibitem[{{Murray} {et~al.}(2018{\natexlab{b}}){Murray}, {Stanimirovi{\'c}},
  {Goss}, {Heiles}, {Dickey}, {Babler}, \& {Kim}}]{murray18}
{Murray}, C.~E., {Stanimirovi{\'c}}, S., {Goss}, W.~M., {et~al.}
  2018{\natexlab{b}}, \apjs, 238, 14

\bibitem[{{Neufeld} {et~al.}(2019){Neufeld}, {DeWitt}, {Lesaffre}, {Cabrit},
  {Gusdorf}, {Tram}, \& {Richter}}]{neufeld19}
{Neufeld}, D.~A., {DeWitt}, C., {Lesaffre}, P., {et~al.} 2019, \apjl, 878, L18

\bibitem[{{Neufeld} {et~al.}(2009){Neufeld}, {Nisini}, {Giannini}, {Melnick},
  {Bergin}, {Yuan}, {Maret}, {Tolls}, {G{\"u}sten}, \& {Kaufman}}]{neufeld09}
{Neufeld}, D.~A., {Nisini}, B., {Giannini}, T., {et~al.} 2009, \apj, 706, 170

\bibitem[{{Odenwald}(1988)}]{odenwald88}
{Odenwald}, S.~F. 1988, \apj, 325, 320

\bibitem[{{Paley} {et~al.}(1991){Paley}, {Low}, {McGraw}, {Cutri}, \&
  {Rix}}]{paley91}
{Paley}, E.~S., {Low}, F.~J., {McGraw}, J.~T., {Cutri}, R.~M., \& {Rix}, H.-W.
  1991, \apj, 376, 335

\bibitem[{{Planck Collaboration} {et~al.}(2011){Planck Collaboration}, {Ade},
  {Aghanim}, {Arnaud}, {Ashdown}, {Aumont}, {Baccigalupi}, {Baker}, {Balbi},
  {Banday}, \& et~al.}]{Planck2011-1.1}
{Planck Collaboration}, {Ade}, P.~A.~R., {Aghanim}, N., {et~al.} 2011, A\&A,
  536, A1

\bibitem[{{Planck Collaboration} {et~al.}(2014{\natexlab{a}}){Planck
  Collaboration}, {Ade}, {Aghanim}, {Armitage-Caplan}, {Arnaud}, {Ashdown},
  {Atrio-Barandela}, {Aumont}, {Baccigalupi}, {Banday}, \&
  et~al.}]{planck2013viii}
---. 2014{\natexlab{a}}, A\&A, 571, A8

\bibitem[{{Planck Collaboration} {et~al.}(2014{\natexlab{b}}){Planck
  Collaboration}, {Abergel}, {Ade}, {Aghanim}, {Alves}, {Aniano}, {Arnaud},
  {Ashdown}, {Aumont}, {Baccigalupi}, {Banday}, {Barreiro}, {Bartlett},
  {Battaner}, {Benabed}, {Benoit-L{\'e}vy}, {Bernard}, {Bersanelli},
  {Bielewicz}, {Bobin}, {Bonaldi}, {Bond}, {Bouchet}, {Boulanger}, {Burigana},
  {Cardoso}, {Catalano}, {Chamballu}, {Chiang}, {Christensen}, {Clements},
  {Colombi}, {Colombo}, {Couchot}, {Crill}, {Cuttaia}, {Danese}, {Davis}, {de
  Bernardis}, {de Rosa}, {de Zotti}, {Delabrouille}, {D{\'e}sert}, {Dickinson},
  {Diego}, {Dole}, {Donzelli}, {Dor{\'e}}, {Douspis}, {Dupac}, {Efstathiou},
  {En{\ss}lin}, {Eriksen}, {Falgarone}, {Finelli}, {Forni}, {Frailis},
  {Franceschi}, {Galeotta}, {Ganga}, {Ghosh}, {Giard}, {Giraud-H{\'e}raud},
  {Gonz{\'a}lez-Nuevo}, {G{\'o}rski}, {Gregorio}, {Gruppuso}, {Guillet},
  {Hansen}, {Harrison}, {Helou}, {Henrot-Versill{\'e}},
  {Hern{\'a}ndez-Monteagudo}, {Herranz}, {Hildebrandt}, {Hivon}, {Hobson},
  {Holmes}, {Hornstrup}, {Hovest}, {Huffenberger}, {Jaffe}, {Jaffe}, {Joncas},
  {Jones}, {Jones}, {Juvela}, {Kalberla}, {Keih{\"a}nen}, {Kerp}, {Keskitalo},
  {Kisner}, {Kneissl}, {Knoche}, {Kunz}, {Kurki-Suonio}, {Lagache},
  {L{\"a}hteenm{\"a}ki}, {Lamarre}, {Lasenby}, {Lawrence}, {Leonardi},
  {Levrier}, {Liguori}, {Lilje}, {Linden-V{\o}rnle}, {L{\'o}pez-Caniego},
  {Lubin}, {Mac{\'{\i}}as-P{\'e}rez}, {Maffei}, {Maino}, {Mandolesi}, {Maris},
  {Marshall}, {Martin}, {Mart{\'{\i}}nez-Gonz{\'a}lez}, {Masi}, {Massardi},
  {Matarrese}, {Mazzotta}, {Melchiorri}, {Mendes}, {Mennella}, {Migliaccio},
  {Mitra}, {Miville-Desch{\^e}nes}, {Moneti}, {Montier}, {Morgante},
  {Mortlock}, {Munshi}, {Murphy}, {Naselsky}, {Nati}, {Natoli}, {Noviello},
  {Novikov}, {Novikov}, {Oxborrow}, {Pagano}, {Pajot}, {Paoletti}, {Pasian},
  {Perdereau}, {Perotto}, {Perrotta}, {Piacentini}, {Piat}, {Pierpaoli},
  {Pietrobon}, {Plaszczynski}, {Pointecouteau}, {Polenta}, {Ponthieu}, {Popa},
  {Pratt}, {Prunet}, {Puget}, {Rachen}, {Reach}, {Rebolo}, {Reinecke},
  {Remazeilles}, {Renault}, {Ricciardi}, {Riller}, {Ristorcelli}, {Rocha},
  {Rosset}, {Roudier}, {Rusholme}, {Sandri}, {Savini}, {Spencer}, {Starck},
  {Sureau}, {Sutton}, {Suur-Uski}, {Sygnet}, {Tauber}, {Terenzi}, {Toffolatti},
  {Tomasi}, {Tristram}, {Tucci}, {Umana}, {Valenziano}, {Valiviita}, {Van
  Tent}, {Verstraete}, {Vielva}, {Villa}, {Wade}, {Wandelt}, {Winkel}, {Yvon},
  {Zacchei}, \& {Zonca}}]{planckXVIIdust}
{Planck Collaboration}, {Abergel}, A., {Ade}, P.~A.~R., {et~al.}
  2014{\natexlab{b}}, \aap, 566, A55

\bibitem[{{Reach} {et~al.}(2017{\natexlab{a}}){Reach}, {Bernard}, {Jarrett}, \&
  {Heiles}}]{reach17b}
{Reach}, W.~T., {Bernard}, J.-P., {Jarrett}, T.~H., \& {Heiles}, C.
  2017{\natexlab{a}}, \apj, 851, 119

\bibitem[{{Reach} {et~al.}(2015){Reach}, {Heiles}, \& {Bernard}}]{reach15}
{Reach}, W.~T., {Heiles}, C., \& {Bernard}, J.-P. 2015, \apj, 811, 118

\bibitem[{{Reach} {et~al.}(2017{\natexlab{b}}){Reach}, {Heiles}, \&
  {Bernard}}]{reach17}
---. 2017{\natexlab{b}}, \apj, 834, 63

\bibitem[{{Reach} {et~al.}(1993){Reach}, {Heiles}, \& {Koo}}]{reach93}
{Reach}, W.~T., {Heiles}, C., \& {Koo}, B.-C. 1993, \apj, 412, 127

\bibitem[{{Reach} {et~al.}(1994){Reach}, {Koo}, \& {Heiles}}]{reach94}
{Reach}, W.~T., {Koo}, B.-C., \& {Heiles}, C. 1994, \apj, 429, 672

\bibitem[{{Reach} {et~al.}(2019){Reach}, {Tram}, {Richter}, {Gusdorf}, \&
  {DeWitt}}]{reach19}
{Reach}, W.~T., {Tram}, L.~N., {Richter}, M., {Gusdorf}, A., \& {DeWitt}, C.
  2019, \apj, 884, 81

\bibitem[{{Stanimirovi{\'c}} \& {Zweibel}(2018)}]{stanimirovic18}
{Stanimirovi{\'c}}, S., \& {Zweibel}, E.~G. 2018, \araa, 56, 489

\bibitem[{{Stark}(1995)}]{stark95}
{Stark}, R. 1995, \aap, 301, 873

\bibitem[{{Tauber} {et~al.}(2010){Tauber}, {Mandolesi}, {Puget}, {Banos},
  {Bersanelli}, {Bouchet}, {Butler}, {Charra}, {Crone}, {Dodsworth}, \&
  et~al.}]{tauber2010a}
{Tauber}, J.~A., {Mandolesi}, N., {Puget}, J.-L., {et~al.} 2010, A\&A, 520, A1

\bibitem[{{Tielens} \& {Hollenbach}(1985)}]{tielens85}
{Tielens}, A.~G.~G.~M., \& {Hollenbach}, D. 1985, \apj, 291, 722

\bibitem[{{van den Bergh} \& {Tammann}(1991)}]{snrateref}
{van den Bergh}, S., \& {Tammann}, G.~A. 1991, \araa, 29, 363

\bibitem[{{Wakker} \& {van Woerden}(1997)}]{wakker97}
{Wakker}, B.~P., \& {van Woerden}, H. 1997, \araa, 35, 217

\bibitem[{{Wolfire} {et~al.}(2003){Wolfire}, {McKee}, {Hollenbach}, \&
  {Tielens}}]{wolfire03}
{Wolfire}, M.~G., {McKee}, C.~F., {Hollenbach}, D., \& {Tielens}, A.~G.~G.~M.
  2003, \apj, 587, 278

\bibitem[{{Wright} {et~al.}(2010){Wright}, {Eisenhardt}, {Mainzer}, {Ressler},
  {Cutri}, {Jarrett}, {Kirkpatrick}, {Padgett}, {McMillan}, {Skrutskie},
  {Stanford}, {Cohen}, {Walker}, {Mather}, {Leisawitz}, {Gautier}, {McLean},
  {Benford}, {Lonsdale}, {Blain}, {Mendez}, {Irace}, {Duval}, {Liu}, {Royer},
  {Heinrichsen}, {Howard}, {Shannon}, {Kendall}, {Walsh}, {Larsen}, {Cardon},
  {Schick}, {Schwalm}, {Abid}, {Fabinsky}, {Naes}, \& {Tsai}}]{wright10}
{Wright}, E.~L., {Eisenhardt}, P.~R.~M., {Mainzer}, A.~K., {et~al.} 2010, \aj,
  140, 1868

\end{thebibliography}

\end{document}